\documentclass{JHEP3}

\usepackage{amsfonts}
\usepackage{amsmath}
\usepackage{amssymb}
\usepackage{latexsym}
\usepackage{graphicx}
\usepackage{cite}

\numberwithin{equation}{section}

\def\fr#1{\mathfrak{#1}}

\def\ZZ{{\mathbb Z}}

\def\IR{{\mathbb R}}

\def\cE{{\cal E}}

\def\cE{\mathcal{E}}

\def\cR{\mathcal{R}}

\newcommand{\be}{\begin{equation}}
\newcommand{\ee}{\end{equation}}
\newcommand{\bea}{\begin{eqnarray}}
\newcommand{\eea}{\end{eqnarray}}
\def\p{\partial }

\def\threeh{{\scriptstyle {3 \over 2}}}
\def\fiveh{{\scriptstyle {5 \over 2}}}
\def\non{\nonumber }

\def\bE{{\bf E}}

\def\hbE{{\hat{\bf E}}}

\def\calE{{\cal E}}

\def\calT{{\mathcal T}}

\def\bs{\backslash}

\def\calF{{\mathcal F}}

\def\calV{{\mathcal V}}

\def\nn{\nonumber} 
\def\half{{\scriptstyle {1 \over 2}}}

\preprint{DAMTP-2010-12, IPHT-T-10/012, IHES/P/10/05, UB-ECM-PF-10/7}

\title{String theory dualities and supergravity divergences}
\author{Michael B. Green\\
 Department of Applied Mathematics and
Theoretical Physics\\
Wilberforce Road, Cambridge CB3 0WA, UK\\
\email{\tt M.B.Green@damtp.cam.ac.uk}}
\author{Jorge G. Russo\\
Instituci{\'o} Catalana de Recerca i Estudis Avan\c cats (ICREA)\\
Departament ECM and Institut de Ciencies del Cosmos, \\
University de Barcelona,  Facultat de Fisica\\
 Av. Diagonal, 647,  Barcelona 08028 SPAIN\\
\email{\tt jrusso@ub.edu}}
\author{Pierre Vanhove\\
Institut des Hautes Etudes Scientifiques\\
Le Bois-Marie, 35 route de Chartres\\
F-91440 Bures-sur-Yvette, France\\
\textrm{and}\\
Institut de Physique Th{\'e}orique,\\
CEA, IPhT, F-91191 Gif-sur-Yvette, France\\
CNRS, URA 2306, F-91191 Gif-sur-Yvette, France\\
\email{pierre.vanhove@cea.fr}}

\abstract{We demonstrate how duality invariance of the low energy expansion of the four-supergraviton amplitude in type II string theory determines the precise  coefficients of multiloop logarithmic ultraviolet divergences of maximal supergravity in various dimensions.  This is illustrated by the explicit moduli-dependence of terms of the form $\partial^{2k}\, \cR^4$, with $k\le 3$, in the effective action. Furthermore,  we show that in the supergravity limit the perturbative contributions are swamped by an accumulation of non-perturbative effects of zero-action instantons. 

}

\begin{document}


\section{Introduction}
\label{sec:intro}
It is well known that string theory provides an ultraviolet completion of supergravity -- there are no ultraviolet divergences in perturbative string theory.   Since perturbative quantum supergravity  arises as the low energy limit of superstring theory it is of interest to see how the field theory ultraviolet divergences appear in the appropriate limit.

 This paper will describe how these field theory divergences are encoded in logarithmic terms that arise in coefficients of the low energy expansion of the type II  superstring four-supergraviton amplitude\footnote{The term ``supergraviton'' refers to the supermultiplet of 256 massless states. The dependence on the helicities of these states arises in the amplitude through a generalised curvature,  $\cR$.} compactified to $D$ dimensions on a $d=(10-D)$-torus, $\calT^d$. These 
 scattering amplitudes have a dependence on the  moduli that is highly constrained by dualities~\cite{Hull:1994ys}, which relate their perturbative and non-perturbative properties.  
 For example,  the  low energy expansion of the  four-supergraviton amplitude generates terms in the effective
 action  of the form  $\partial^{2k}\,  \cR^4$ ,  where  $\cR^4$ is  a
 specific contraction of four generalised curvature tensors, which depends on the superhelicites and momenta of the external states..   The coefficients of such terms are functions of the moduli that are
 invariant  under  discrete  symmetries  associated  with  the  groups
 $E_{d+1(d+1)}(\ZZ)$ (which are discrete  versions of real split forms
 of the corresponding Lie groups of rank
 $d+1$)\footnote{For $d\leq 5$ $E_{1(1)}(\IR)=SL(2,\IR)$, $E_{2(2)}(\IR)=
   SL(2,\IR)\times  GL(1)$, $E_{3(3)}(\IR)=SL(3,\IR)\times SL(2,\IR)$,
   $E_{4(4)}(\IR)=SL(5,\IR)$,  $E_{5(5)}(\IR)=  SO(5,5,\IR)$  and  for
   $5\leq d\leq 7$  $E_{d+1(d+1)}$ is the real split form of the
     exceptional Lie group $E_{d+1}$.}, and contain the complete perturbative and non-perturbative information about the amplitude.    In contrast to string theory, classical maximal  supergravity   is invariant
   under the  continuous version of  these groups, which  implies that
   the Feynman rules are independent of the moduli. As a consequence, perturbative supergravity amplitudes do not depend on the moduli.  
  However, this ignores the presence of infinite towers of non-perturbative charged BPS black hole states, which probably invalidates the use of the perturbative approximation, whether or not there are ultraviolet divergences, as we will argue later.

In   a   recent    paper~\cite{Green:2010wi}, which will be summarised in section~\ref{sec:reviewsec},   we   determined   the
non-perturbative expressions for the coefficients of a number of terms
in  the  low  energy  expansion  of  the  four-supergraviton  amplitude  of
maximally supersymmetric string theory compactified from 10 dimensions
to $D= (10-d)$ dimensions on  a $d$-torus.  The most detailed analysis
was for  the analytic  part of the  amplitude with $d\le  3$, although
certain features of  the nonanalytic terms and the $3< d \leq 7$
cases were  also determined.  The simplest  interactions considered in
some   detail   in~\cite{Green:2010wi}   (extending   earlier   work
in~\cite{Green:1997tv,Green:1997di,Green:1997as,Green:1999pu,Green:1998by,Green:2005ba,Green:2008bf,Russo:1997mk,Kiritsis:1997em,Pioline,Lambert,Basu:2007ru,Basu:2007ck},
see also recent discussions in~\cite{Pioline:2010kb,LambertNew})
were  $\cR^4$ and  $\partial^4\, \cR^4$,  for which  the  coefficients are
special  combinations  of Eisenstein  series  of  the kind  considered
in~\cite{Langlands}.   The  coefficient   of  the  $\partial^6\,  \cR^4$
interaction    coefficient    is    a   more    general    automorphic
function~\cite{Green:2005ba,Green:2008bf}. A  thorough analysis of these
coefficients demonstrated that they  reduce to the correct expressions
in  three  different  limits:  (i) String  perturbation  theory;  (ii)
Decompactification  from $D$  to  $D+1$ dimensions  when  a radius  of
$\calT^d$ becomes  large; (iii) The  semi-classical eleven-dimensional
supergravity  limit,  in  which  the  M-theory  torus,  $\calT^{d+1}$,
becomes large and loop calculations in eleven-dimensional supergravity
are valid.  It was also  argued that in certain `critical' dimensions,
$D_L$,  the leading logarithmic ultra-violet  divergences of  $L$-loop maximal
supergravity  are reproduced\footnote{The  critical  dimension at  $L$
  loops is  the lowest (possibly  non-integer) dimension in  which the
  theory     has    ultraviolet     divergences.}. As    remarked
in~\cite{Green:2010wi},  particular examples  of such  behaviour arise
for the  $\cR^4$ interaction with $(D_1 =8,\  L=1)$, the $\partial^4\,
\cR^4$ interaction  with $(D_2=7,\ L=2)$ and  the $\partial^6\, \cR^4$
interaction    with    $(D_3=     6,    \    L=3)$.     The  structure of the coefficients determined in~\cite{Green:2010wi}        will        be        reviewed        in
section~\ref{sec:reviewsec}. 
 
In the following  section we will present a detailed argument that the logarithmic factors that arise in the automorphic coefficients of the string theory higher derivative interactions  indeed determine the values of logarithmic ultraviolet divergences in loop amplitudes of maximal supergravity.  To be precise, we will see in section~\ref{sec:sugralogs} that the logarithmic terms in the  coefficients of 
$\partial^{2k}\, \cR^4$ interactions with $k=0$ in $D=8$, $k=2$ in  $D=7$, and $k=3$ in $D=6$ are equal to the logarithmic terms that arise in maximal supergravity after subtracting the ultraviolet divergences.
The    $\partial^6\,    \cR^4$    coefficient  function was    not    determined
in~\cite{Green:2010wi} and  so, for completeness, it  will be obtained
in  appendix~\ref{sec:d6r4}.

 In  addition, there  are `non-leading'  logarithmic
terms that  arise in dimensions  $D> D_L$, which are identified
with   further   logarithmic   ultraviolet  divergences   in   maximal
supergravity.  For example, there  is a single pole, $1/\epsilon$, and
a double-pole,  $1/\epsilon^2$, in dimensionally  regularised two-loop
maximal   supergravity  in  $D=8$   dimensions  that   contributes  to
$\partial^6\cR^4$ (whereas the $D_3=6$ single pole contributes to $\partial^6 \cR^4$).  
Another new feature arises in the field theory since  the one-loop $\cR^4$ divergence requires a counterterm.  This contributes to a one-loop `triangle' diagram in which one vertex is the counterterm, which results in another $1/\epsilon^2$ contribution~\cite{Green:1999pu}, which  we will also evaluate in section~\ref{sec:sugralogs}.  The sum of these contributions gives rise to $\log$ and $\log^2$ terms that are reproduced by the string theory coefficient of this interaction. 
In order to compare the field theory and string theory expressions it is important to use consistent normalisation conventions, which are briefly outlined in appendix~\ref{sec:normalisations}.

 In section~\ref{sec:sugrainst} a connection will be made 
with the issue  of whether quantum supergravity might  be a consistent
theory that  can be  obtained as a  decoupling limit  of closed-string
string  theory, much  as ${\cal  N}=4$ super  Yang--Mills in four
dimensions can be obtained as a decoupling limit of open string theory.  
It was pointed out in~\cite{Green:2007zzb} that this is probably  far from the case even if the individual terms of the perturbative expansion are finite.   The problem is due to the presence of infinite towers of non-perturbative states, which correspond in toroidally compactified  string theory to massive Kaluza--Klein modes, winding modes, Kaluza--Klein monopoles and wrapped $p$-branes of various kinds. It was shown in~\cite{Green:2007zzb} that the
supergravity limit is  one in which towers of  states becomes massless
and  the restriction  of  the spectrum  to  the massless  perturbative
states --  the basic assumption in  supergravity -- is  not a sensible
approximation   to  the   theory.  In an analogous fashion the  simple examples in this paper involve a condensation of zero-action instantons, as will be demonstrated in section~\ref{sec:sugrainst}, based on the explicit expressions for the coefficients of the $\cR^4$ and $\partial^4\, \cR^4$ interactions.

Although the complete structure of the automorphic coefficient functions has not been determined beyond order $\partial^6 \cR^4$, a certain amount is known about higher order terms based on analysis of one and two loop amplitudes in eleven-dimensional supergravity compactified to $D=9$ nine dimensions on $\calT^2$ in~\cite{Green:2008bf}.  This will be used as the basis of a speculative discussion in section~\ref{sec:comments}  suggesting that the $\partial^8\cR^4$ interaction is not protected by supersymmetry against perturbative corrections at genus five and higher, which would have significant implications for the onset of ultraviolet divergences in perturbative maximal supergravity.

 The paper will end with a  short discussion of these results in section~\ref{s:summary}.

 \section{Summary of duality invariant coefficients in the low energy expansion}
 \label{sec:reviewsec}

In~\cite{Green:2010wi} we were concerned with properties of the low-momentum expansion of the four-supergraviton amplitude. It is useful to separate the 
$D$-dimensional amplitude into the sum of analytic and non-analytic terms,
 \be
 A_D(s,t,u)= A_D^{analytic}(s,t,u) + A_D^{nonan}(s,t,u)\,,
 \label{ampsplit}
 \ee
 where the analytic part has a low energy expansion in powers of the Mandelstam variables ($s=-(k_1+k_2)^2$, $t=-(k_1+k_4)^2$, $u=-(k_1+k_3)^2$)  of the form
  \be
 A_D^{analytic} = \sum_{p=0}^\infty\sum_{q=-1}^\infty\cE^{(D)}_{(p,q)}(\phi_{K\bs G})\, \sigma_2^p\, \sigma_3^q\, \cR^4 \,.
 \label{analytic}
 \ee
 This   is   the  general  symmetric  polynomial in the Mandelstam invariants, which enter in the dimensionless combinations
 \be
 \label{sigdef}
 \sigma_n = (s^n + t^n + u^n)\, {\ell_D^{2n}\over 4^n}\,,
 \ee
where   $\ell_D$  is  the   Planck  length   in  $D$   dimensions.  
 The
 coefficient functions, $\cE^{(D)}_{(p,q)}(\phi_{K\bs G})$,  are  functions  of  the
 symmetric  space moduli,  $\phi_{K\bs G}$, which are  the scalar fields, of the coset space $K\bs
 G$  appropriate to compactification  on a $d=(10-D)$-torus (where  $G$  is   $E_{d+1(d+1)}(\IR)$  and $K$ is its  maximal compact subgroup).
 They   are required to be automorphic functions that are
 invariant  under the $D$-dimensional  duality group,  $E_{d+1(d+1)}(\mathbb Z)$,
 The  expansion is one  in which $k_i\cdot  k_j\, r^2 \ll  1$ and
 $k_i\cdot  k_j\,\ell_D^2 \ll  1$,  where  $r$ is  any  radius of  the
 toroidal dimensions,  $\ell_D$ is the  $D$-dimensional Planck length,
 and $k_i$ and $k_j$ are any of the external momenta.  The nonanalytic term, $A^{nonan}_D$, contains singularities due to thresholds in which internal lines of the perturbative contributions to the amplitude are on-shell.  
  The separation of the amplitude into the two pieces in~\eqref{ampsplit} is well defined at low orders in the low-energy expansion, where there are few perturbative contributions to the amplitude.
 
 It is convenient to express the analytic part of the amplitude in terms of a local one-particle irreducible effective action,
 \be
 \label{effactgen}
 S_D^{local} = \sum_{p\geq0,q\geq-1}  \ell_D^{8+2k-D}\,\int d^Dx\, \sqrt{-G^{(D)}}\,\calE^{(D)}_{(p,q)}\,  \partial^{2k} \cR^4\,
 \ee
where  $k=2p+3q$ and $G^{(D)}$  is the  determinant of  the space-time
metric in the Einstein frame.

\subsection{Constraints on the coefficients}
\label{s:conscoeff}

It is  clear that maximal supersymmetry imposes  strong constraints on
the  structure of the  coefficient functions.   In particular,  it was
shown in~\cite{Green:1998by} that  type IIB supersymmetry requires the
coefficient of the $\cR^4$ interaction  in ten dimensions to satisfy a
Laplace eigenvalue equation (with  a particular eigenvalue), for  which the  unique  solution compatible with  string perturbation theory is a nonholomorphic  Eisenstein series,
$\calE^{(10)}_{(0,0)} (\Omega)= \bE_{\threeh}(\Omega)$, where $\Omega$
is the  complex modulus of the IIB  theory.  So far there  has been no
progress in  generalising this supersymmetry argument  to higher order
interactions   (see,   however,~\cite{Sinha:2002zr,Basu:2008cf})  or   higher-rank
groups, but the following  indirect arguments (given in~\cite{Green:2010wi})  lead to appropriate generalised Laplace eigenvalue equations satisfied by the coefficient functions in the compactified theory.
It was argued in~\cite{Green:2010wi} that in the decompactification
limit $r_{10-D}/\ell_{D+1}\to\infty$ the Laplace operator, $\Delta^{(D)}$, on $K\bs G$   becomes
\begin{equation}
 \Delta^{(D)}\to\Delta^{(D+1)}+{D-2\over 2(D-1)} \,
 (r_{10-D}\partial_{       r_{10-D}})^2+{D^2-3D-58\over      2(D-1)}\,
 r_{10-D}\partial_{ r_{10-D}}\, ,
\end{equation}
and the eigenvalues $\lambda_{(p,q)}^{(D)}$ of the interaction coefficients
$\cE^{(D)}_{(p,q)}$ satisfy the equation
\begin{equation}
 \lambda_{(p,q)}^{(D)}    -\lambda_{(p,q)}^{(D+1)}=   {2p+3(q+1)\over
   (D-1)(D-2)}\, (D^2-3D-52+4p+6q)\, .
\end{equation}
Using  the   ten   dimensional  values   $\lambda^{(10)}_{(0,0)}=3/4$,
$\lambda^{(10)}_{(1,0)}=15/4$      and      $\lambda^{(10)}_{(0,1)}=12$
determined
in~\cite{Green:1997tv,Green:1998by,Green:1999pu,Sinha:2002zr,Green:2005ba},       we
deduce that the coefficients of the terms discussed in~\cite{Green:2010wi}  satisfy the following set of Laplace eigenvalue equations with source terms,
\begin{eqnarray}
  \left( \Delta^{(D)}                  -{3(11-D)                  (D-8)\over
    D-2}\right)\,\cE^{(D)}_{(0,0)}&=&  6\pi\,\delta_{D-8,0} \,,
\label{laplaceeigenone}\\   
\left( \Delta^{(D)}  -{5(12-D) (D-7)\over D-2}\right)\,\cE^{(D)}_{(1,0)}&=&40\zeta(2)\, \delta_{D-7,0} \,,
\label{laplaceeigentwo}\\   
\left( \Delta^{(D)} -{6(14-D) (D-6)\over D-2} \right)\,\cE^{(D)}_{(0,1)}&=&-\left(\cE_{(0,0)}^{(D)}\right)^2 + 120\zeta(3)\,  \delta_{D-6,0}\,,
\label{laplaceeigenthree}
\end{eqnarray}
where     the    coefficient   of the $\delta_{D-6,0}$     in
equation~(\ref{laplaceeigenthree}), which was not determined
in~\cite{Green:2010wi}, is derived in appendix~\ref{sec:d6r4}.
 Although most of the discussion in~\cite{Green:2010wi} focused on explicit solutions of these equations with 
$7\le  D\le  10$,  the  iterative  argument  linking
dimensions $D$ and  $D+1$ shows that they hold  more generally for all
dimensions $D\geq3$.

The structure of equations~\eqref{laplaceeigenone} and~\eqref{laplaceeigentwo}  generalizes  the Laplace equation satisfied by the  $\cR^4$ interaction in $D=10$ dimensions~\cite{Green:1997tv}.   A notable feature of these eigenvalue equations is the presence of the Kronecker delta sources  which are non-zero in the dimensions in which the eigenvalues vanish.  These are the critical dimensions, $D_L$, which are the lowest dimensions in which $L$-loop maximal supergravity has ultraviolet divergences.    Equation~\eqref{laplaceeigenthree}, satisfied by the coefficient of the $\partial^6 \, \cR^4$ interaction, has a source term  that is quadratic in the coefficient of the $\cR^4$ interaction, which  can also be interpreted to be  a consequence of maximal supersymmetry~\cite{Green:2005ba}.  In addition the Kronecker delta contributes in $D_3 =6$ dimensions, which is  again the dimension in which the  eigenvalue vanishes and is also the lowest dimension in which $L=3$ supergravity has an ultraviolet divergence.  Interactions of higher order will not be discussed here in any detail.  However, some of their properties in $D=9$ dimensions were determined in~\cite{Green:2008bf}, which indicated that the coefficients are {\it sums} of automorphic functions that satisfy equations that are generalisations of~\eqref{laplaceeigenthree}.

 \begin{figure}[ht]
 \centering\includegraphics[width=8cm]{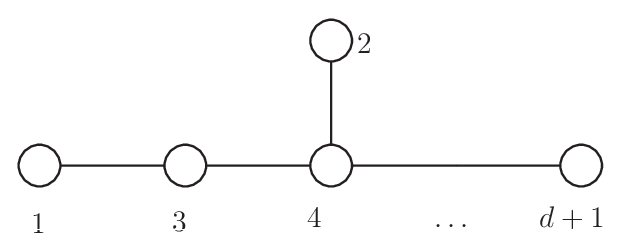}
 \caption{\label{fig:dynkin}  The  Dynkin  diagrams of  the  U-duality
   groups $E_{d+1(d+1)}$ ($0\leq d\leq 7$)}  
 \end{figure}

The solutions of~\eqref{laplaceeigenone}--\eqref{laplaceeigenthree} are highly constrained by imposing boundary conditions that require them to reproduce known features of string/M theory in various limits.  These limits are: 
\begin{description}
\item[(i)]
The  limit in which  one radius,  $r_d$, of  the string  theory torus,
$\calT^d$, becomes  large, $r_d\gg  \ell_{D+1}$ so that  the amplitude
effectively      decompactifies       from      $D=10-d$      to      $D+1$
dimensions.\footnote{This  limit is equivalent  to $r_d\gg
  \ell_s$  with the $D+1$-dimensional  string coupling  $y_{D+1}$ held
  fixed.} Since the external  momenta, $k_i$ ($i=1,2,3,4$), are fixed,
this is a limit in which $k_i\cdot k_j\, r_d^2\gg 1$, which lies outside the
range of  validity of  the original expansion.   In order for  the low
energy expansion to be valid  in $D+1$ dimensions it is necessary that
$k_i\cdot k_j\, \ell_{D+1}^2\ll 1$.
Although  this interchange of  limits might  generally be  expected to
pose problems, it  does not at low orders  in the derivative expansion
that are  considered here  because only a  finite number of  powers of
$r_d$    occur.    To   be    precise,   the    $\cR^4$   coefficient,
$\calE^{(D)}_{(0,0)}$  has  two  distinct   powers  of  $r_d$  in  its
expansion, so (ignoring coefficients) the expansion has the form
 \be
\left(\ell_{D}\over\ell_{D+1}\right)^{8-D}\,\cE^{(D)}_{(0,0)}\to {r_d\over\ell_{D+1}}\,
\cE^{(D+1)}_{(0,0)}+ 
   \left({r_d\over \ell_{D+1}}\right)^{8-D} \,.
   \label{rfourseries}
 \ee
 The term that grows linearly  with $r_d$ gives the finite contribution to the $\cR^4$ interaction  in the large $r_d/\ell_{D+1}$ limit.  The second term is the $n=1$ term of an infinite series  of the schematic form $r_d^{8-D}\, (s\, r_d^2)^n\, \cR^4$, which resums in a manner that converts the first nonanalytic threshold of the $D$-dimensional amplitude to that of the  $(D+1)$-dimensional amplitude.   For simplicity, we have suppressed a $\log r_d/\ell_{D+1}$ factor that multiplies the second term when $D=7$ and $D=8$.

The $\partial^4\, \cR^4$ coefficient, $\calE^{(D)}_{(1,0)}$, has three power-behaved 
terms in its expansion,
\be
\left(\ell_{D}\over\ell_{D+1}\right)^{12-D}\,\cE^{(D)}_{(1,0)}\to  {r_d\over\ell_{D+1}}\,
\cE^{(D+1)}_{(1,0)} + \left(r_d\over    \ell_{D+1}\right)^{12-D}+   \left({r_d\over    \ell_{D+1}}\right)^{6-D}   
\cE^{(D+1)}_{(0,0)}\,.
\label{d4r4series}
\ee
Again the  term linear in $r_d$  gives the finite  contribution to the
interaction  in   the  large-$r_d$   limit,  while  the   second  term
contributes the $n=2$ term  of the series $r_d^{8-D}\, (s\, r_d^2)^n\,
\cR^4$ that resums to give  the first nonanalytic threshold.  The last
term  contributes the  first term  of  a second  infinite series  that
resums to  give the second  $(D+1)$-dimensional nonanalytic threshold.
We have suppressed a $\log(r_5/\ell_{6})$ factor multiplying the third
term when $D=5$ and $D=6$.

The $\partial^6\, \cR^4$ coefficient, $\calE^{(D)}_{(0,1)}$, has four
terms in its expansion
\begin{equation}\begin{split}
\left(\ell_{D}\over\ell_{D+1}\right)^{14-D}\,\cE^{(D)}_{(0,1)}&\to  {r_d\over\ell_{D+1}}\,
\cE^{(D+1)}_{(0,1)} + \left(r_d\over    \ell_{D+1}\right)^{14-D}+   \left({r_d\over    \ell_{D+1}}\right)^{8-D}   
\cE^{(D+1)}_{(0,0)}\cr
&+    \left({r_d\over    \ell_{D+1}}\right)^{4-D}   
\cE^{(D+1)}_{(1,0)}+\left(r_d\over    \ell_{D+1}\right)^{15-2D}\,.
\label{d6r4series}
\end{split}\end{equation}
The term  linear in $r_d$ again  gives the finite  contribution to the
interaction in the large-$r_d$  limit, the second term contributes the
$n=3$ term  of the  series that resums  to give the  first nonanalytic
threshold and the  third term contributes a second  term to the series
that sums  to the second  threshold.  The fourth term  contributes the
first term  to a  new infinite  series that resums  to give  the third
$(D+1)$-dimensional nonanalytic threshold.   The last
term is  the $n=1$  term of the  series that  resums to give  a second
nonanalytic supergravity threshold contribution.
Again, we have ignored logarithmic factors that arise for $D=3$ and
$D=7$.

 \item[(ii)]The  limit of  string  perturbation theory.   This is  the
   limit in  which the $D$-dimensional string  coupling becomes small,
   so that  $y_D = g_s^2\,  \ell_s^d/V_{(d)} \ll 1$, where  $V_{(d)} =
   r_1r_2  \dots  r_d$   is  the  volume  of  $\calT^d$   and  $g_s $  is  the  string   coupling.   In  this  limit  each
   coefficient possesses a finite set  of terms that are power behaved
   in $y_D$.  In string frame a term of order $y_D^{-1+h}$ corresponds
   to a term of genus-$h$ in closed string perturbation theory.  In addition there is an infinite set of exponentially suppressed instanton contributions. A great deal is known about the low-energy expansion of the four-supergraviton amplitude directly from string perturbation theory at genus-one and genus-two, and a certain amount at genus-three.
 
 \item[(iii)]  The limit in  which the  M-theory torus  becomes large,
   $\calV_{(d+1)}\gg  \ell_{11}^{d+1}$.   In   this  limit,  $r_d  \gg
   \ell_{11}$, with $k_i\cdot k_j\ell_{11}^2\ll1$ the semi-classical (Feynman diagram) approximation to eleven-dimensional supergravity is expected to be a good approximation.   A variety of calculations in compactified eleven-dimensional supergravity at one loop and two loops provide much information about this limit~\cite{Green:1997as,Russo:1997mk,Green:1999pu,Green:2006gt,Green:2008bf}. 
  
\end{description}

 In each of these three cases a specific  parameter becomes large and there is a finite number of terms that are power-behaved in this parameter, together with an infinite series of exponentially suppressed terms. 
The sum of  power behaved  terms contributes the zero  Fourier mode,  or `constant'
term  with  respect to  the  angular  parameters  that enter  in  the
off-diagonal entries of the matrix  $N$ (the unipotent radical) of the
standard Levi decomposition of  a maximal parabolic subgroup of $G$, $P=MN$, where $M$ is the Levi factor for the corresponding subgroup of $G$.   Such constant terms are are obtained  by deleting  specific  nodes of  the $E_{d+1(d+1)}$  groups.
Numbering    the     $E_{d+1(d+1)}$ nodes as indicated in
figure~\ref{fig:dynkin},  in  limit~(i)  node  $d+1$  is  deleted,  in
limit~(ii) node  1 is deleted, and  in limit~(iii) node  2 is deleted.
The   exponentially   suppressed  terms   in   each   case  have   the
interpretation of BPS instanton  contributions due to D-instantons and
a   variety   of  wrapped   euclidean   $p$-branes.   Although   these
contributions have not been  analysed in detail they should correspond
to  1/2-BPS  states  in  the  $\cR^4$  case,  1/4-BPS  states  in  the
$\partial^4\cR^4$  case, and 1/8-BPS  states in  the $\partial^6\cR^4$
case (see for example~\cite{Pioline:2010kb}  for a recent viewpoint of
such contributions in the 1/2- and 1/4-BPS cases).
 A novel feature appears in the $\partial^6\cR^4$ case, where D-instanton/anti D-instanton pairs with zero net instanton number arise, giving exponentially suppressed contributions to the constant terms.
 
 The coefficient functions discussed in~\cite{Green:2010wi} are in precise  agreement with all the boundary data in these three limits and also satisfy the Laplace equations in~\eqref{laplaceeigenone}--\eqref{laplaceeigenthree}.  
 In  the case of  the $\cR^4$  and $\partial^4\cR^4$  interactions the
 solutions are combinations of  Eisenstein series for the rank-$(d+1)$
 duality groups.  In the case of the $\partial^6\cR^4$ interaction the
 solution is  a less familiar  automorphic function. Although  we have
 not  proved that  these solutions  are  unique, given  the number  of
 conditions that need to be satisfied it seems unlikely that there are
 ambiguities (although we cannot rule out the possibility of cusp forms).  We will briefly review the kinds of series that enter into the solutions (more details are given in appendix B of 
~\cite{Green:2010wi}).

\subsection{Definition and properties of Eisenstein series.}

The `minimal parabolic' Eisenstein series for a group $G$ is defined with respect to a complex vector $\lambda$ in the weight space of the Lie algebra  $\fr g$ as~\cite{Langlands}
 \begin{equation}
  \label{e:minparab}
  \bE^G_\lambda(g)= \sum_{\gamma\in G(\mathbb Q)\bs  B(\mathbb  Q) }
  e^{\langle \lambda+\rho, H(g\gamma )\rangle}\,,
\end{equation}
 where  $\rho$ is half the sum of the positive roots, $\langle  \cdot,\cdot\rangle$ is  the inner
product  on the  root system  of  $G$, $H(g)$ is a vector in the Cartan subalgebra, and  $B$ is a Borel subgroup of $G$.   These Eisenstein series  are eigenfunctions of the invariant differential operators of $K\bs G$.  In 
particular,    they    are    eigenfunctions      of    the
Laplacian,\footnote{
Invariance under
  $K$ implies that the eigenvalue  of the Laplacian is the same as
  the  value of the  second-order Casimir  of
  $G$. }
\begin{equation}
  \label{e:LaplaEisenstein}
  \Delta_{K\bs                  G}                 \,\bE^G_\lambda(g)=
  2(\langle\lambda,\lambda\rangle-\langle\rho,\rho\rangle)\,\bE^G_\lambda(g)\,.
\end{equation}
Whereas  the $SL(2)$  Eisenstein series  depends on  a  single complex
parameter  $s$, for higher-rank  groups there  are $r  ={\rm rank}(G)$
such  parameters,  $s_k$  ($k=1,\dots,r$),  that are  related  to  the
entries in $\lambda$. 
The minimal parabolic Eisenstein series has a poles for various values
of $\lambda$~\cite{Langlands}, but
the special cases of interest to us are ones that are obtained by taking the multiple residue on the poles at  $s_k=0$ for all $k \ne \alpha$,  so only $s\equiv s_\alpha$ is non-zero, where $\alpha$ is a particular node of the Dynkin diagram of $G$.  In other words 
we set 
\bea
\lambda_{d-\alpha+1} - \lambda_{d-\alpha}-1 &=&2s\,, \nn\\
\lambda_{d-k+1}-\lambda_{d-k}-1 &=& 0\,, \qquad {\rm all}\ 1\le k\ne \alpha\le d-1\,.
\label{sldeis}
\eea
This defines the {\it maximal} parabolic Eisenstein series for a particular parabolic subgroup of $G$ associated with the Dynkin label
$[0^{\alpha-1}\, 1\, 0^{r-\alpha}]$, which will be denoted by\footnote{The conventional $SL(2)$ Eisenstein series will be denoted by
$\bE_s \equiv \bE^{SL(2)}_{[1];s}$.} $\bE^{(G)}_{[0^{\alpha-1}\, 1\, 0^{r-\alpha}];s}$.

These Eisenstein series can be expressed as sums over integer lattices.  In the simplest cases these sums can be analysed directly.  For example, the $SL(n)$ series $\bE^{SL(n)}_{[0^{\alpha-1}\, 1\, 0^{n-\alpha-1}];s}$ is given by
\be
\bE^{SL(n)}_{[0^{\alpha-1}\,      1\,      0^{n-\alpha-1}];s}      =
{\sum_{\{m_i\}\in \ZZ^n}}^\prime
 {1\over (d^{[i_1\dots i_\alpha]}\, g_{i_1 j_1}\dots g_{i_\alpha j_\alpha}\, d^{[j_1\dots j_\alpha]})^s}\,,
\label{slneisen}
\ee
where $g_{ij}$  ($i,j=1, \dots, n$) is an  $SL(n)$ matrix parametrizing
the coset $SO(n)\bs SL(n)$, $d^{[i_1
  \dots i_\alpha]}$ is the antisymmetrized product of $\alpha$ integer
vectors,   $d^{[i_1   \dots   i_\alpha]}=  m_1^{[i_1}m_2^{i_2}   \dots
m_\alpha^{i_\alpha]}$  and the sum  excludes the  values at  which the
denominator vanishes. 

 However, for other duality groups these lattice
sums  are  more  subtle.  This  is  illustrated  by  the case  of  the
$SO(d,d)$ series\footnote{ The $d=5$ case is of relevance as the $D=6$ U-duality
  group  $SO(5,5)$, which also arises as the duality symmetry of perturbative string theory 
 in $D=5$  with a different interpretation of the
  moduli.} $\bE^{SO(d,d)}_{[1,0^{d-1}];s}$, which has the representation (motivated by the expression for one-loop perturbative amplitude for string theory compactified on $\calT^d$~\cite{Green:2010wi})
\be
\bE^{SO(d,d)}_{[1\,0^{d-1}];s}=   {\pi^{s}\over 2\zeta(2s+2-d)\Gamma(s)}\,\int_{\calF_{SL(2,\mathbb Z)}} {d^2\tau\over
     \tau_2^2} \bE_{s+1-{d\over2}}(\tau)\,(\Gamma_{(d,d)}(\tau)-V_{(d)})\,,
\label{so55}
\ee
where  $\Gamma_{(d,d)}(\tau)$  is  the  standard  lattice  factor  for
compactification  of  the  one-loop  string  amplitude  on  $\calT^d$,
$V_{(d)}$  is the volume  of $\calT^d$  and the  integral is  over the
fundamental domain of $SL(2,\ZZ)$.  
  The corresponding representations
of the other  $SO(d,d)$ series, as well as  the $E_{6(6)}$, $E_{7(7)}$
and  $E_{8(8)}$  series  have  not  been  determined  (as  far  as  we
know). However,  it is possible to  analyse all the  series from their
definition~\eqref{e:minparab}.   This procedure  has been  carried out
and will  be reported in  detail elsewhere. 

The arguments of~\cite{Green:2010wi} (and earlier work reviewed therein) lead to the  $\cR^4$ coefficients that enter the Einstein-frame action~\eqref{effactgen}\footnote{10B indicates the ten-dimensional type IIB theory.},
\be
\begin{tabular}{llclclcll} 
$D=10$ &$\calE^{(10B)}_{(0,0)}= \bE_{\threeh}(\Omega)$ &   $d=0$  \,,\\
 $D=9$ &$\calE_{(0,0)}^{(9)}  =\bE_{3\over2}(\Omega)\,\nu_1^{-{3\over7}} +4\zeta(2)\,\,\nu_1^{{4\over7}}$ &  $d=1$\,, \\
  $D=8$ &$\calE^{(8)}_{(0,0)} = \lim_{\epsilon\to 0} \left(\bE^{SL(3)}_{[10];\threeh + \epsilon}  +2 \bE_{1-2\epsilon}(U) \right)  = \hbE^{SL(3)}_{[10];\threeh}  +2 \hbE_1(U) $ & $d=2$\,,\\
 $3\leq D < 8$  &$ \calE^{(D)}_{(0,0)} = \bE^{E_{d+1(d+1)}}_{[10\cdots0];\threeh} $&
 $2<d \leq 7$\,.\\
 \end{tabular}
  \label{tabr4}
  \ee
   Each  Eisenstein series  in these  equations is  a function  of the
   moduli that parametrize  the coset space $K\bs G$  of the U-duality
   group $G=E_{d+1(d+1)}$ by its maximal compact subgroup $K$.  In the following we will omit the arguments of the Eisenstein series unless this is likely to lead to confusion.  The quantity $\nu_1$ is defined in terms of the radius of the circular dimension in the type IIB theory, $r_B$, by $\nu_1= (r_B/\ell_{10})^2$. 
The individual series  in the third line have poles  at $\epsilon = 0$
but these  poles cancel  in their sum.  The symbol $\hbE$  indicates a
series  that  is regularised  by  subtracting  a  pole in  $\epsilon$.
In~\cite{Green:2010wi} it has been explicitly verified that these coefficients satisfy all the required boundary conditions, as well as the Laplace eigenvalue equations~\eqref{laplaceeigenone} for $D\ge 6$ (and is extended to $D\le 5$ in a forthcoming paper in collaboration with Stephen Miller~\cite{GMRV}).

The coefficients of the $\partial^4 \cR^4$ interactions in dimensions $7\le D \le 10$ are given by 
 \be
 \begin{tabular}{llclclcll}
 $D=10$ & $\calE^{(10B)}_{(1,0)}(\Omega) =\frac12\, \bE_{\fiveh}(\Omega)$ &$d=0$\\
  $D=9$ &$\calE^{(9)}_{(1,0)} =\frac12\, \nu_1^{-{5\over 7}}\,\bE_{5\over2}(\Omega)+{2\zeta(2)\over15}\, \nu_1^{{9\over7}} \,
 \bE_{3\over2}(\Omega)+  {4\zeta(2)\zeta(3)\over15}\, \nu_1^{-{12\over7}}$ & $d=1$ \\
$D=8$ &  $  \mathcal{E}^{(8)}_{(1,0)}=\frac12\,\bE^{SL(3)}_{[10];\fiveh}
 -4\, \bE^{SL(3)}_{[10];-\frac12}  \,\bE_2(U) $ & $d=2$ \\
$D=7$ & $\cE_{(1,0)}^{(7)} = \lim_{\epsilon\to 0}\left(\frac12\,\bE^{SL(5)}_{[1000];\fiveh+\epsilon }+{3\over \pi^3}\bE^{SL(5)}_{[0010];\fiveh-\epsilon}\right)
= \frac12\, \hbE^{SL(5)}_{[1000];\frac52}+{3\over \pi^3}\hbE^{SL(5)}_{[0010];\fiveh}$ & $d=3$ \\
 \end{tabular}
 \label{tabd4r4}
\ee
The poles in the last line again cancel,  yielding a  finite
expression.  These expressions satisfy  all the boundary conditions in
the  three  degeneration limits  described  earlier,  as  well as  the
Laplace eigenvalue equations~\eqref{laplaceeigentwo}.
The extension of  these expression for $D\leq 6$  will be presented in
reference~\cite{GMRV}. 

The  solutions of  the inhomogeneous  equations for  the coefficients,
$\calE^{(D)}_{(0,1)}$, of the  $\partial^6 \cR^4$ interaction are more
complicated  and given  in~\cite{Green:2010wi}  for $D  \ge 7$.   Some
details  of the  $D=6$ case  are presented  in appendix~\ref{sec:d6r4}
since it is of particular interest to this paper.

\section{Logarithmic terms and ultraviolet divergences in supergravity} 
\label{sec:sugralogs}

One of the intriguing features of the expressions for the coefficients
in~\cite{Green:2010wi}  is the manner  in which  potential divergences
cancel.   The  Eisenstein series  that  enter  into the  coefficients,
$\calE^{(D)}_{(0,0)}$,            $\calE^{(D)}_{(1,0)}$            and
$\calE^{(D)}_{(0,1)}$,  have singularities at  specific values  of the
parameter $s$.   This reflects the pole  at $s=1$ in  the Riemann zeta
function, $\zeta(s)$.  However, the precise combinations of Eisenstein
series that enter  are ones for which the  pole residues cancel.  This
is a  manifestation of the consistency of  string perturbation theory.
Although  the  poles  cancel,   there  are  residual  terms  that  are
logarithms of a modulus, which are important elements in the structure
of the  amplitude.  We will here  focus on logarithms  of the coupling
constant,  $\log y_D$.   These enter  in  cases where  the low  energy
supergravity  limit has  a logarithmic ultraviolet  divergence, manifested as a
pole in dimensional regularisation. 

These logarithmic terms are the origin of the Kronecker delta terms on
the                           right-hand                          side
of~\eqref{laplaceeigenone}--\eqref{laplaceeigenthree}.          Roughly
speaking this follows from the  fact that part of the Laplace operator
acting on $\calE^{(D)}_{(p,q)}$ contains $y^2 \partial^2_y\, \log y =-1$.  The
simplest  example   of  this  phenomenon   is  seen  in   the  $\cR^4$
coefficient,     $\calE^{(8)}_{(0,0)}$     in     $D=8$     dimensions
in~\eqref{tabr4},   the   next  being   in   the  $\partial^4   \cR^4$
coefficient,          $\calE^{(7)}_{(1,0)}$          in          $D=7$
dimensions~\eqref{tabd4r4}.      The    third    example     is    the
$\partial^6\,\cR^4$   coefficient  in   $D=6$  dimensions,   which  is
presented in appendix~\ref{sec:d6r4}.

\subsection{Logarithmic thresholds in  the Einstein frame}

Closed   string   perturbation  theory   is   an   expansion  in   the
$D$-dimensional  coupling constant,  in  which the  genus-$h$ term  is
proportional to  $y_D^{h-1}$ when evaluated in the  string frame.  The
four-supergraviton amplitude  contains terms  that are non-analytic  in the
Mandelstam invariants  due to massless thresholds  that are determined
by unitarity.  Up to the order in the low energy expansion that we are
considering in  this paper these are  the same thresholds  as those of
maximal  supergravity where  they  arise at  $L$  loops in  dimensions
$D_L=4+6/L$~\cite{Green:2006gt}.   In the  string amplitude  these are
schematically of the form, 
\be
\ell_s^{8-D_L}\, y_{D_L}^{L-1}(\ell_s^2\, s)^{n_L}\, h_L(x)\, \cR^4\, \log (-\ell_s^2\, s\, f_L(x))\,,\qquad n_1=0\,, \  n_2=2\,, \ n_3=3\,,
\label{schemrfour}
\ee
where $f_L$  and $h_L$ are complicated functions  of the dimensionless
variable $x=  -t/s = 1 +u/s$, the  details of which do  not concern us
(see~\cite{Green:2006gt}     for     a     discussion     of     these
contributions).  The power of $\ell_s$  in the overall factor is fixed
by the  power of  the Mandelstam invariants  and the  dimension $D_L$.
Importantly,  apart from the  explicit power  of the  string coupling,
$y_{D_L}$, there is no dependence  on the moduli in the overall factor
multiplying these nonanalytic terms,  although $f_L(x)$ does depend on
the moduli other than $y_{D_L}$. Transforming from the string frame to the Einstein frame is equivalent
to replacing $\ell_s$ by $\ell_D$ using $\ell_D^{D-2} = \ell_s^{D-2}\,
y_D$. This implies that the Mandelstam invariants are rescaled so that 
\be
\ell_s^2\, s =  y_D^{-\frac{2}{D-2}}\, \ell_D^2\,  s, \qquad {\rm or} \qquad \log (-\ell_s^2\, s) = \log (-\ell_D^2 \, s) - \frac{2}{D-2}\, \log y_D\,.
\label{rescale}
\ee
The contribution  to the amplitude  in~\eqref{schemrfour} is therefore
equal to the Einstein frame expression
\be
\ell_{D_L}^{8-D_L}\,  (\ell_{D_L}^2\,s)^{n_L}\, h_L(x)\, \cR^4\, \left(\log (-\ell_{D_L}^2s f_L(x))-{2\over D_L-2}\, \log y_{D_L}\right)
\label{schemrfourein}
\ee
So  we  see that  when  the  Mandelstam  invariants are  expressed  in
Einstein frame units  the non-analytic $\log s$ term  in the amplitude
leads to a  term proportional to $\log y_D$ in  the analytic part.  In
this discussion there is an  ambiguity in the scale of the logarithms,
but this does not affect the overall coefficient and is independent of
the moduli, so for our purposes it can be ignored. In other words, the
coefficient of the  $\log y_D$ term in Einstein  frame is $-2/(D_L-2)$
times the coefficient of the $\log(- \ell_D^2s)$ term.

On the other hand, in supergravity the factor of $\log s$ arises as an
infrared   threshold   singularity   accompanied  by   a   logarithmic
ultraviolet  divergence.   If  this  is regulated  by  an  ultraviolet
momentum  cutoff  $\Lambda$,  it  results   in  a  term  of  the  form
$\log(-s/\Lambda^2)$, where  the $\log  \Lambda$ can be  subtracted by
addition of  a local  counterterm.  In dimensional  regularisation the
ultraviolet divergence appears as an $\epsilon$ pole in the amplitude
evaluated in $D=D_L+2\epsilon$ dimensions. The logarithm appears after
subtracting   the   pole    and   using   $\lim_{\epsilon\to   0}
((-s/\mu)^{2\epsilon}-1)/\epsilon  \sim \log(-s/\mu)$, where  $\mu$ is
an  arbitrary scale.  Needless to  say, since  the coefficient  of the
$\log$  is  determined  by  unitarity  it  is  not  sensitive  to  the
regularisation scheme adopted.

The  conclusion  is that  the  logarithmic  terms  in the  automorphic
functions,  determine the  coefficients  of the  $\log  s$ factors  in
$A^{nonan}_D$,  and  hence  the  logarithmic terms  that  represent  the
ultraviolet  divergences  in  supergravity.   The  following  examples
illustrate this feature of the amplitudes in the three cases, $D_1=8$,
$D_2=7$ and  $D_3=6$.  The conventions used to  compare the amplitudes
in    string    theory    and    supergravity   are    exhibited    in
appendix~\ref{sec:normalisations}.

\medskip
\noindent
{\bf $\bullet$ The $\cR^4$ interaction in $D= 8$ dimensions}
\smallskip

It  was shown in~\cite{Kiritsis:1997em,Green:2010wi} that the coefficient $\calE^{(8)}_{(0,0)}$ in~\eqref{tabr4} has the  perturbative expansion
 \be
 \calE^{(8)}_{(0,0)}  = {2\zeta(3)\over y_8}+2(\hbE_{1}(T)+\hbE_{1}(U))+{2\pi\over3}\log y_8+O(e^{-(y_8T_2)^{-\half}},e^{-(\,y_8/T_2)^{-\half}})\,.
  \label{e:E322reg}
  \ee
  In   this   case   there   is   no   overall   power   of   $\ell_8$
  in~\eqref{effactgen} so  this expression is also  the coefficient in
  the  string frame and  the power-behaved  terms are  identified with
  tree-level  ($h=0$) and  genus-one  ($h=1$) contributions,  together
  with the $\log y_8$ term.  The latter is a signal of a genus-one $\log (-s\, \ell_s^2)$ term in the string frame,  where there can be no $\log y_8$, as argued above.

This expression  can be  compared with the  expression that  arises in
dimensionally    regularised   one-loop   maximal    supergravity   in
$D=8+2\epsilon$,  where  the $\epsilon$  pole  is  associated with  an
ultraviolet divergence.  The field theory amplitude   given in~\cite{Green:1982sw} 
is
\be
A_{R}^{tree} +A_{R^4}^{1-loop} \propto \cR^4\,\left(\,{64\over stu\,
    \ell_8^6} +\ \hat I_1(\ell_8^2 k_i\cdot k_j)\right)\,,
\label{rfoureps}
\ee
where we have included the tree-level term proportional to $\cR^4/stu$
in  order  to  display   the  relative  normalisations  (we  refer  to
appendix~\ref{sec:normalisations} for details) and 
\be
\label{stusum}
\hat I_1(\ell_8^2
k_i\cdot k_j)=I_1(s,t)+I_1(t,s)+I_1(s,u)+I_1(u,s)+I_1(t,u)+I_1(u,t)\,,
\ee
 with 
 \be
 I_1(\ell_8^2k_i\cdot k_j) = \lim_{\epsilon\to 0} (I_1^\epsilon(\ell_8^2
k_i\cdot k_j) + {2\pi \over \epsilon})\,,
 \label{polesub}
 \ee
 and 
\begin{equation}
  I^\epsilon_1(s,t)={2\pi\over 3}\left( {1\over \epsilon} +\log\left(-{\ell_8^2 t\over\mu}\right)\right)\,\int_0^1 dx {t (1-x)\over sx-t(1-x)}+{2\pi \over 3}\int_0^1 dx
  {t (1-x)\,\log(1-x)\over sx-t(1-x)}+O(\epsilon)\,
\label{epspole}
\end{equation}
($\mu$ is an arbitrary constant).  It is easy to see that this expression contains a logarithmic term.  Summing over the terms in~\eqref{stusum}  and rescaling the metric to the string frame  using the identity $\ell_8^2 =y_8^{1/3}\, \ell_s^2$  gives 
 \be
A^{1-loop}_{R^4}(\ell_s^2 k_i\cdot k_j) = A^{1-loop}_{R^4}(\ell_8^2 k_i\cdot k_j) + \frac{2\pi}{3}\log y_8\,\cR^4 \,.
\label{rfournonan}
\ee 
Therefore,   the  ${2\over 3}\pi \log  y_8$  contribution in the   coefficient $\calE^{(8)}_{(0,0)}$ in ~\eqref{e:E322reg} implies the presence of the threshold logarithm, which is given in supergravity by the dimensionally regularised expression $\hat I_1(\ell_8^2 k_i\cdot k_j)$.  
 So the coefficient of the logarithmic ultraviolet divergence associated with the field theory pole in 
 \eqref{epspole} is precisely the coefficient of the $\log y_8$ required by U-duality.

\medskip
\noindent
{\bf $\bullet$ The $\partial^4 \cR^4$ interaction in $D=7$ dimensions}
\smallskip

  The coefficient of this interaction is  $\ell_7^5\, \calE^{(7)}_{(1,0)}$  which is defined by~\eqref{tabd4r4} and  was shown in~\cite{Green:2010wi} to have the small-$y_7$  the expansion
   \begin{equation}\begin{split}
  \cE_{(1,0)}^{(7)} & =  {\zeta(5)\over y^2_7}+{3\over\pi^3 y_7}\,
    \bE^{SL(4)}_{[010];\frac52}
+                {2\over3}
(\hbE^{SL(4)}_{[100];2}+\hbE^{SL(4)}_{[001];2})+{8\pi^2\over15}\log y_7 \cr
&+ O(e^{-(y_7v_3)^{-\half}},e^{-(y_7\ell_s/r_i)^{-\half}})\,,
\label{e:pert7d}
\end{split}\end{equation}
where $v_3=(r_1r_2r_3)/\ell_s^3$. The various powers of $y_7$ in 
this expression  correspond to tree-level  ($h=0$), genus-one ($h=1$)
and genus-two  ($h=2$) terms.  This is seen by transforming to the string frame where  the terms  are of
order $y_7^{-1+h}$ using the fact that $\ell_7^5  = \ell_s^5\,  y_7$. The logarithmic  term here implies the existence
of  a   genus-two  threshold  term   of  the  form   $4\pi^2/3\,  y_7
\log(-s\ell_s^2)$  in string frame using \eqref{rescale} again. 
 
We can compare the coefficient of the $\log y_7$ term in \eqref{e:pert7d} with the ultraviolet divergence of two-loop maximal supergravity in $D=7$ dimensions,   which was evaluated using dimensional regularisation in~\cite{Bern:1998ug} and gave (once again including the tree-level term in  order to  compare  normalisations),
\be
A_{R}^{tree}+A^{2-loop}_{\partial^4R^4}\propto\cR^4\, \ell_7\,
\left({64\over  stu\,  \ell_7^6}   +
\hat  I_2(\ell_7^2k_i\cdot k_j)\right)\,,
\label{d4r4eps}
\ee
where the regularised two-loop contribution is defined by
 \begin{equation}
\hat I_2(\ell_7^2k_i\cdot k_j)  = \lim_{\epsilon\to  0}  (I_2^{\epsilon}  +{\ell_7^4
   \over 2 \epsilon}\,{\pi^2\over12}\,(s^2+t^2+u^2) )
\,,
 \label{polesubd4}
 \end{equation}
with 
\be
\label{twoeps}
I_2^\epsilon(\ell_7^2\,k_i\cdot k_j)=(\ell_7^2\, s)^2\,
(I^{P\, \epsilon}_2(s,t)+ I^{P\, \epsilon}_2(s,u)+ I^{NP\, \epsilon}_2(s,t)+  I_2^{NP\, \epsilon}(s,u))+{\rm perms}(s,t,u)\,.
\ee
Here  $P$ and  $NP$  denote contributions  from  planar and  nonplanar
double-box  Feynman  integrals,  which  are  defined  via  dimensional
regularisation in  $D=7+2\epsilon$ dimensions using equation~(4.3) of~\cite{Bern:1998ug}
where 
\begin{equation}\label{I2:gen}
  I_2^{X\, \epsilon}(s,t)
  =2^{D-7}\pi^{2D-12}\,\Gamma(7-D)\,(-\ell_D^2s)^{D-7}
  \,\int_0^1d^7\nu\delta(1-\sum_{i=1}^7\nu_i)\,\Delta_X^{{14-3D\over2}}+ \cdots\,,
\end{equation}
where $X=P$ or $NP$ and $\dots$ indicates terms that do not contribute to the logarithm and $\Delta_X$ is given by~\cite{Bern:1998ug}
\begin{eqnarray} 
 \Delta_P  &=& (\nu_1+\nu_2+\nu_3)(\nu_4+\nu_5+\nu_6) + \nu_7(1-\nu_7)\,,\nn \\
\Delta_{NP} &=& (\nu_1+\nu_2)(\nu_3+\nu_4) + (\nu_1+\nu_2+\nu_3+\nu_4)(\nu_5+\nu_6+\nu_7) \,. 
 \label{deltadef}
   \end{eqnarray}
Expanding~(\ref{I2:gen}) one gets (see appendix~C of~\cite{Bern:1998ug})
\begin{eqnarray}
 \hat I_2
  &=&\ell_7^4\,{\pi^2\over12}        \,
 (s^2\log(-{\ell_7^2s\over\mu})+ t^2\log(-{\ell_7^2t\over\mu})+u^2\log(-{\ell_7^2u\over\mu}))+\cdots
\end{eqnarray}
Substituting in~\eqref{d4r4eps} and using $\ell_7^2=\ell_s^2\, y_7^{2/5}$ gives the relation
\begin{equation}
 A^{2-loop}_{\p^4R^4}(\ell_7^2k_i\cdot  k_j)= y_7\, A^{2-loop}_{\p^4R^4}(\ell_s^2k_i\cdot  k_j)+
  {8\pi^2\over15}\,\log y_7\,\sigma_2\,\ell_7\, \cR^4\, ,
 \label{dfourfournanon}
\end{equation}
which shows that $L=2$ supergravity produces a string-frame genus-two threshold logarithm together with a $\log y_7$ term that is identical to the one contained in the automorphic coefficient function $\calE^{(7)}_{(1,0)}$ in~\eqref{e:pert7d}.   In other words, as with the $\cR^4$ interaction, we can identify the precise coefficient of the logarithm associated with an $\epsilon$ pole  in dimensional regularisation of two-loop maximal supergravity in $D=7+2\epsilon$ dimensions with the coefficient of the logarithm in the duality-invariant coefficient, $\calE^{(7)}_{(1,0)}$. 

\medskip
\noindent
{\bf $\bullet$ The $\partial^6 \cR^4$ interaction in $D=6$ dimensions}
\smallskip

In this case the  coefficient, $\calE^{(6)}_{(0,1)}$ is an
automorphic function for the  U-duality group $SO(5,5)$ that satisfies
the   inhomogeneous   equation~\eqref{laplaceeigenthree},  which   has
vanishing  eigenvalue but non-zero Kronecker delta term when $D=6$.    The  solution  of  this   equation  is  less
straightforward than  the earlier cases.   Since this case  was hardly
discussed   in~\cite{Green:2010wi}   (whereas  the   $\partial^6\cR^4$
coefficients          for         $D>6$          were         obtained
in~\cite{Green:2005ba,Basu:2007ck,Green:2010wi}),   a   discussion  is
included in  the appendix, from which we  see that
the coefficient $\cE^{(6)}_{(0,1)}$ has the perturbative expansion 
\begin{equation}\begin{split}
\cE^{(6)}_{(0,1)}       &={2\zeta(3)^2\over  3\,     y_6^3}+       {1\over
  y_6^2}\,({2\zeta(3)\over3}\,\bE^{SO(4,4)}_{[1000];1}+{8\zeta(4)\over 69\pi} \bE^{SO(4,4)}_{[1000];4})+ {1\over
    y_6}\, F_2^{SO(4,4)} \cr
&+          {4\zeta(2)\over 105}        \,
  (\hbE^{SO(4,4)}_{[0001];3}+\hbE^{SO(4,4)}_{[0010];3})+
  15\zeta(3)\,\log y_6+n.p.\,.
\label{d6r4coupling}
\end{split}\end{equation}
where $n.p.$ stands for various non perturbative contributions
evaluated in appendix~\ref{sec:d6r4} where the function $ F_2^{SO(4,4)}$ is also discussed.  In this case the powers of the string coupling, $y_6$, correspond to tree-level, genus-one, genus-two and genus-three.  The three-loop  contribution involves  the regularized
$SO(4,4)$         series        $\hbE^{SO(4,4)}_{[0001];3}$        and
$\hbE^{SO(4,4)}_{[0010];3}$.
  In  particular, the  logarithmic term  is  a sign  of a  genus-three
  logarithm  associated  with  a  term  in the  string  frame  of  the
  schematic form  $y_6^2\zeta(3) \log(-\ell_s^2 s)$.
  
  Once again this can be compared with dimensionally regularised supergravity, which has a three-loop contribution to  the $\partial^6\cR^4$
  amplitude in $D=6$ dimensions of the  form (again adding in the  tree-level amplitude to
  compare normalisations) 
  \be
A_{R}^{tree}+A^{3-loop}_{\partial^6R^4}\propto
\cR^4\,\ell_6^2\,\left({64\over  stu\,  \ell_6^6}  +\hat I_3(\ell_6^2\,k_i\cdot k_j)\right)\,.
\label{d6r4eps}
\ee
The function $\hat I_3$   is   a   sum   of   many
contributions~\cite{Bern:2007hh,Bern:2008pv} that is given by 
using equation~(5.12) of~\cite{Bern:2007hh}, which gives 
\begin{equation}\begin{split}
\hat                                           I_3&=\lim_{\epsilon\to0}
(I_3^\epsilon+{5\zeta(3)\over3\epsilon}\,{\sigma_3\over4^3})\cr
&={1\over4^3}\,({1\over9}       (t^3+u^3-2s^3)-\zeta(3)      \,(t^3+u^3+3s^3)  )    
\log(-\ell_6^2\,s/\mu)\cr
&+{1\over4^3}\, ({1\over9}  (s^3+u^3-2t^3)-\zeta(3)      \,(s^3+u^3+3t^3)      )
\log(-\ell_6^2\,t/\mu)\cr
&+{1\over4^3}\,( {1\over9}  (s^3+t^3-2u^3)-\zeta(3)      \,(s^3+t^3+3u^3)      )
\log(-\ell_6^2\,u/\mu)+\cdots\,.
\end{split}\end{equation}
The  expression for  $\hat I_3$  can be  deduced  from equation~(5.19)
of~\cite{Bern:2007hh}   (using   the    $D=7$   two-loop   result   in
equation~(5.14) to establish normalisations). 

Substituting this expression into~\eqref{d6r4eps} leads to the transformation of the three-loop amplitude from Einstein frame  to the  string  frame (using the relation $\ell_6^2=\ell_s^2\, y_6^{1/2}$),
\begin{equation}
 A^{3-loop}_{\p^6R^4}(\ell_6^2\,k_i\cdot     k_j)= y_6^2 A^{3-loop}_{\p^6R^4}(\ell_s^2\,k_i\cdot    k_j)+
   {5\over2}\zeta(3)\,\log    y_6    \,\sigma_3\,\ell_6^2\,\cR^4\,.
\end{equation}
Therefore, the coefficient of the $\log y_6$ term in $\calE^{(6)}_{(0,1)}$ in~\eqref{d6r4coupling} determines 
the  coefficient   of  the  logarithmic  terms   associated  with  the
$\epsilon$ pole. The  relative factor of 6 between  the coefficient of
the  $\log  y_6$  in  this  expression  and  in  $\calE^{(6)}_{(0,1)}$
\eqref{d6r4coupling} is   a puzzle that is presumably
due to difficulties in comparing the normalisations in the two computations (since the coefficient of the $\log s$ factor is fixed by unitarity the results should surely  be equal).

\medskip
\noindent
{\bf $\bullet$ The $\partial^6 \cR^4$ interaction in $D=8$ dimensions}
\smallskip

The examples  discussed so  far are ones  in the  critical dimensions,
$D_L=4+6/L$,  for $L=1,2,3$.   There are,  however,  other ultraviolet
logarithms that arise in dimensions $D>D_L$ for any value of $L$.  The
simplest of these appears to arise in the one-loop in ten dimensions,
where  there  is  a  threshold  that  is  schematically  of  the  form
$s\,\ell_{10}^2\, \log(-s\, \ell_{10}^2)\, \cR^4 + {\rm perm}(s,t,u)$.
However, under the rescaling $\ell_{10}^2 = \ell_s^2\, y_{10}^{1/4}$ 
the shift is $(s+t+u)\, \log y_{10} = 0$, so the logarithmic term vanishes.

 The simplest  nontrivial example is  the two-loop amplitude  in $D=8$
 dimensions,  which   has  both  $\log$   and  $\log^2$  divergences
 associated with a single and double pole  multiplying $\partial^6\, \cR^4$ in dimensional regularisation in $D=8+2\epsilon$ dimensions.
The presence of these supergravity divergences is again encoded in the duality invariant $\partial^6\, \cR^4$ coefficient function, $\calE^{(8)}_{(0,1)}$,  which satisfies \eqref{laplaceeigenthree} with $D=8$.   In this case the source term on the right-hand side of \eqref{laplaceeigenthree} is the square of the $\cR^4$ coefficient, $\calE^{(8)}_{(0,0)}$, which itself has a one-loop $\log y_8$, as exhibited in \eqref{e:E322reg}.
The solution of this equation  has the perturbative expansion given in
equation~(5.20) in~\cite{Green:2010wi}, which has the logarithmic terms,
 \begin{equation}
      \cE^{(8)}_{(0,1)}=\cdots+
{\pi\over 9}  ({\pi\over6}+\bE^{pert}_{(0,0)}) \log y_8 - {\pi^2\over 27} \log^2 y_8 \Big)+n.p.\,,
\label{xpert}
\end{equation}
where  $\bE^{pert}_{(0,0)}$ is  the perturbative  part of  the $\cR^4$
interaction which has the expansion  given in~\eqref{e:E322reg}.  The term in~\eqref{xpert} involving the tree-level part of    $\bE^{pert}_{(0,0)}$  is a stringy threshold effect that was
discussed  in \cite{Green:2008uj}.  It  contains the  factorisation of
the string loop into the product  of a tree-level $\cR^4$ factor and a
massless    pole    factor.      It is noteable that the one-loop  part of
$\bE^{pert}_{(0,0)}$ gives a contribution $2\pi^2\log y_8/27$   in \eqref{xpert}, which flips the sign of the  explicit  $-\pi^2\log^2   y_8^2/27$  term.  

In this case the corresponding  $D=8$ supergravity field theory calculation involves the sum of two kinds of diagrams: (i)  The two-loop diagrams evaluated in~\cite{Bern:1998ug}.  (ii) A contribution involving the $\cR^4$ counterterm that cancels the one-loop divergence  - it is necessary to include the diagram in which this counterterm,  is inserted as a vertex in a one-loop diagram.

In the first of these contributions, (i),
the  double $\epsilon$  pole of  the dimensionally  regulated two-loop
amplitude   of  maximal   supergravity   in~\eqref{I2:gen}  leads to a $\log^2 s$
term   
\begin{equation}\begin{split}
\hat           I^{(i)}_2&=\lim_{\epsilon\to0}           (I_2^\epsilon+\ell_8^6\,(-{\pi^2\over192\epsilon^2}-{43\pi^2\over3456\epsilon})(s^3+t^3+u^3)\cr
&-\ell_8^6\,{\pi^2\over96\epsilon}(s^3 \log(-{s\ell_8^2\over \mu})+t^3 \log(-{t\ell_8^2\over \mu})+u^3 \log(-{u\ell_8^2\over \mu}))\,.
\end{split}\end{equation}
The  $\log^2  y_8$ term  should correspond  to  the
double-pole in  $\epsilon$ in  the two-loop supergravity  amplitude in
$D=8+2\epsilon$~\cite{Bern:1998ug}. 

 However, in eight dimensions the complete amplitude also includes contribution (ii)
due to the one-loop $\cR^4$ counterterm, which has an $\epsilon$ pole,
inserted into a one-loop diagram.  This results in a triangle diagram,
which makes an essential additional contribution,  $I_2^{(ii)}$, to the  $\partial^6\cR^4$ ultraviolet divergence in eight dimensions.
Its  overall normalisation is  difficult to  determine, but its value can be fixed
by requiring that its contribution cancels the $\log(-\ell_8^2 s/\mu)/\epsilon$ pole
in  $I_2^\epsilon$.\footnote{We would  like to  thank Hugh  Osborn for
  discussions on this issue.} Although this is not a completely independent check of the normalisation (unlike the previous cases), it shows the precise origin of the different structures that contribute to give the string theory result.  With this proviso, the counterterm contribution is 
\begin{equation}
  I_2^{(ii)\,         \epsilon}         ={16\over (4\pi)^3}\,{\pi\over\epsilon}\,      \ell_8^4\,  (s^2
  \,I_\triangleright^\epsilon(s)+t^2\,
  I_\triangleright^\epsilon(t)+u^2\, I_\triangleright^\epsilon(u))\, ,
\end{equation}
where  
\begin{equation}
 I^\epsilon_\triangleright(s)=                \int                 {d^D\ell\over
   \ell^2(\ell-k_1)^2(\ell-k_1-k_2)^2} 
={2^{5-D}\pi^{D+3\over2}\over
 (D-4)}\,{(-\ell_D^2 s)^{D/2-3}\over\Gamma\left(D-3\over2\right)\sin(\pi(3-{D\over2}))}\,
 \label{eq:triangle}
\end{equation}
 in which one vertex is the  $\cR^4$ counterterm and in which the loop
 integral generates a second power of $1/\epsilon$ when evaluated in $D=8+2\epsilon$. As a result this
 contribution has the form 
\begin{equation}
\hat I_2^{(ii)} = \lim_{\epsilon\to 0}\Big(I_2^{(ii)\,\epsilon} -
\ell_8^6\,{\pi^2\over96}\,
   {s^3+t^3+u^3\over\epsilon^2}
+    \ell_8^6\,{\pi^2\over96}\,           {s^3\log(-
     {\ell_8^2s\over\mu})+t^3\log(-{\ell_8^2                  t\over\mu})+u^3\log(-{\ell_8^2
     u\over\mu})\over\epsilon}\Big)
\label{trianglediag}
\end{equation}
which gives another contribution to the double pole.  It is striking that the addition of the $1/\epsilon^2$ term arising in the counterterm diagram, $I^{(ii)}_2$,  flips the sign of the $1/\epsilon^2$ term in $I^{(i)}_2$.    This corresponds precisely to the  flip of the sign of the total $\log^2 y_8$ term due to the addition of the one-loop sub-divergence in~\eqref{xpert}.

The total contribution obtained by adding (i) and (ii) is given by
\begin{equation}\begin{split}
    \hat      I_2^{(i)}+ \hat I_2^{(ii)} &=-\ell_8^6\,\frac{\pi^2}{192}      (s^3     \log^2(-{s\ell_8^2\over
      \mu})+t^3     \log^2(-{t\ell_8^2\over
      \mu})+u^3     \log^2(-{u\ell_8^2\over
      \mu}))\cr
&-\ell_8^6\,{5\pi^2\over144}\, (s^3     \log(-{s\ell_8^2\over
      \mu})+t^3     \log(-{t\ell_8^2\over
      \mu})+u^3     \log(-{u\ell_8^2\over
      \mu})) \, .
\end{split}\end{equation}
Substituting this in
\begin{equation}
  A_{R}^{tree}+A^{2-loop}_{\partial^6R^4}\propto\cR^4\, 
\left({64\over  stu\,  \ell_8^6}   +
\hat  I_2(\ell_8^2k_i\cdot k_j)\right)\,,
\label{d6r4epss}
\end{equation}
  the amplitude transforms as (using $\ell_8^2=\ell_s^2\, y_8^{1/3}$) 
\begin{equation}
A^{2-loop}_{\p^6R^4}(\ell_8^2k_i\cdot k_j)=y_8\,A^{2-loop}_{\p^6R^4}(\ell_s^2k_i\cdot k_j)-{\pi^2\over 27}\log^2 y_8 \, \sigma_3\,\cR^4+\cdots\,,
\end{equation} 
where $\cdots$ denotes terms with a single power of $\log y_8$.  So we see that 
there is agreement between the coefficient of the $\log^2 y_8$ term in
the automorphic coefficient $\calE^{(8)}_{(0,1)}$ and the supergravity
calculation.   As  is evident  from~\eqref{xpert},  the string  theory
coefficient automatically includes the term with the single logarithm,
$\bE^{pert}_{(0,0)}\,\log y_8$, which corresponds to the one-loop term
that has to be subtracted in the field theory calculation in~\cite{Bern:1998ug}.

\section{The supergravity limit and instanton corrections}
\label{sec:sugrainst}

We turn  now to consider  the particular  low energy  limit of string  theory that
should relate to perturbative quantum supergravity in $D$ dimensions,  which is an
expansion in powers of $k_i\cdot k_j\, \ell_D^2 \ll 1$, where the $D$-dimensional Planck length is fixed while $\ell_s\to 0$, so the string excitation masses become large.
  Since
\be
\ell_D^{D-2} = y_D\, \ell_s^{D-2}\,,
\label{scalesrelate}
\ee
it follows that the limit of interest is one in which the $D$-dimensional string coupling becomes large, 
\be
\lim_{\ell_s\to 0} y_D = \frac{g_s^2\, \ell_s^d}{r_1\cdots r_d}\to \infty\,.
\label{limony}
\ee
In addition, in order to arrive at the the field theory limit in which there is a single massless supermultiplet, the masses of all other massive states must become large and decouple.   This requires, in particular,  $r_i \to 0$ so that non-zero Kaluza--Klein masses are large, and $\ell_s^2/r_i \to 0$ for the winding masses to become large.

\subsection{The perturbative terms}

In the  $y_D\to\infty$ limit the perturbative  term with
the  highest  power  of  $y_D$ dominates the others.
For $D> D_L=4+6/L$ this is a positive power of $y_D$ so the leading term diverges, signifying  a  power-behaved
divergence in supergravity.  The simplest example of this is in $D=10$
string theory, where the genus-one term corresponds, in this limit, to
a   term   of   the   form  $\ell_s^{-2}   \cR^4   =   y_{10}^{1/4}
\ell_{10}^{-2}\,  \cR^4$. This diverges  in the  large-$y_{10}$ limit,
signifying  the   quadratic  divergence   of  the  one-loop   term  in
supergravity. 

When  $D= D_L$ the dominant {\it perturbative} term in the $y_D\to \infty$ limit is the $\log y_{D_L}$ term,   which gives  the supergravity  logarithm    for  each of the three interactions described in
equations~\eqref{e:E322reg},~\eqref{e:pert7d},~\eqref{d6r4coupling}. 

 For $D<D_L$ the  perturbative terms vanish in the field
theory limit since they involve  inverse powers of $y_D$ that arise in
the translation from string frame  to Einstein frame.  This is clearly
seen from  the specific examples  of the $\cR^4$ interaction  in $D=7$
and $D=6$ dimensions, as follows.

\noindent  $\bullet$ The $\cR^4$ interaction in $D= 7$ dimensions has perturbative terms that are given 
by~\cite{Green:2010wi},
 \begin{equation}\label{e:R47dpert}
\cE^{(7)}_{(0,0)}\bigg|_{pert}=    y_7^{-\frac{1}{5}}\,\left({2\zeta(3)\over y_7}+2\pi \, \bE^{SO(3,3)}_{[100];\frac12} \right)\,,
 \end{equation}
where the factor of $y_7^{-1/5}$ comes from the relation $ \ell_7=\ell_s\, y_7^{1/5}$ in converting from string units to Planck units in seven dimensions.

\noindent
$\bullet$ The $\cR^4$ interaction in $D= 6$ dimensions has the perturbative terms~\cite{Green:2010wi},
 \begin{equation}
 \cE^{(6)}_{(0,0)} \bigg|_{pert} = y_6^{-\half}\, \left({2\zeta(3)\over
    y_6}+ 2\bE^{SO(4,4)}_{[1000];1}\right)\,,
   \label{perturbinsix}
 \end{equation}
where the factor $y_6^{-1/2}$ again arises from the conversion from string frame to Einstein frame (using $\ell_6= \ell_s\, y_6^{1/4}$).

In both these  examples the perturbative terms vanish  in the $y_D \to
\infty$ limit, which is a  statement of the well-known fact that there
is no local $\cR^4$ interaction in maximal supergravity for $D<8$.  In
these dimensions  the leading contribution beyond  the tree-level term
is a  non-local interaction  roughly of the  form $s^{\frac{D-8}{2}}\,
\cR^4$     (although     its      precise     details     are     more
complicated~\cite{Green:1982sw}).  A  similar argument shows  that the
perturbative   parts  of   the   $\partial^4\,  \cR^4$   coefficients,
$\calE_{(1,0)}^{(D)}$, vanish in the  $y_D\to \infty$ limit for $D<7$.
The  same  is  true  for $\calE^{(D)}_{(0,1)}$  when  $D<6$.   Whether
analogous  statements  apply  to   higher  orders  in  the  derivative
expansion has not been demonstrated.

However, there  are important  non-perturbative effects in  the string
amplitude         that          swamp         the         perturbative
contribution~\cite{Green:2007zzb} as will be demonstrated next.

\subsection{Supergravity limit including the instanton terms}

Nonperturbative  effects  are,  of  course,
suppressed in string perturbation theory,  in which $y_D$ is small and
other moduli  are fixed.  However, the $y_D\to  \infty$ limit produces
an infinite series  of instanton terms with actions  that become small
in  the   limit  under  consideration.   For   example,  consider  the
exponential   terms   in   the  expansion   of   $\calE^{(8)}_{(0,0)}$
in~\eqref{e:E322reg},  which correspond to  a series  of $D$-instanton
terms (with action $(y_8\, T_2)^{-1/2 }$) and of wrapped $D$-string
instanton terms (with action $(y_8/T_2)^{-1/2}$).  Although these are
both suppressed when $y_8$ is small, at least one of these series
is  unsuppressed for large $y_8$.   This is  an instanton
manifestation of  the effect described  in~\cite{Green:2007zzb}, where
it was shown that in dimensions $D>3$ there are necessarily towers of
non-perturbative   particle  states  that   become  massless   in  the
supergravity  limit. This will  now be  demonstrated in  our explicit
examples.

\bigskip
{\bf $\bullet$ The $\cR^4$ interaction in $D=8$}
\medskip

In this case we will reexamine the exact expression for  $\cE_{(0,0)}^{(8)} = \hat \bE^{SL(3)}_{[10];{3\over 2}} +2 \hat \bE^{SL(2)}_{[1];1}(U)
$ in~\eqref{tabr4}  in the limit $y_8\to \infty $.  
Consider first the expansion of $ \bE^{SL(3)}_{[10];s} $ in the limit $y_8\to \infty $, which is defined by\cite{Kiritsis:1997em,Green:2010wi}
 \begin{equation}\label{eEmX}
  \bE^{SL(3)}_{[10];s}=\sum_{(m_1,m_2,m_3)\neq (0,0,0)} \, {y_8^{s\over 3}\over \left( y_8  \Big(m_1+m_2\Omega_1+ m_3(B_{RR} +\Omega_1 T_1 ) \Big)^2
+{|m_2+m_3 T |^2\over T_2}\right)^s}\,.
 \end{equation}
The limit $y_8\to \infty $ can be studied by separating the leading piece, which is the term with $m_1=0$ in~\eqref{eEmX},  and then perform Poisson resummations.
This    expansion    is   analogous    to    the    one   in    (B.52)
in~\cite{Green:2010wi}, but with the substitution $(\nu_2,\Omega)\to (y_8,T)$.
For  $s\neq 3/2$ this gives
 \begin{eqnarray}
   \label{ocho}
\nn \bE^{SL(3)}_{[10];s} &=&    y_8^{{s\over3}}    \,      \bE_s(T)+2
   \pi\,{\Gamma(s-1)\over\Gamma(s)} \,\zeta(2s-2) y_8^{3-2s\over
       3}\\
 &+&
   {2\pi^s\over   \Gamma(s)}\,    y_8^{3-s\over6}\,   T^{1-s\over2}_2\,
 \sum_{m_1,m_2\atop          m_3\neq0}\,         \left|m_2-m_1T\over
   m_3\right|^{s-1}\\
\nn&\times&K_{s-1}\left(2\pi|m_3|\sqrt{{m_2^2\over     T_2}+m_1^2T_2}\,
  y_8^{1/2}\right)\, e^{2i\pi m_3(m_1 B_{\rm RR}+m_2 B_{\rm NS})}\,.
 \end{eqnarray}
{}Regularising the pole at $s=3/2$ gives
\begin{eqnarray}
  \label{e:E32reg}
\hbE^{SL(3)}_{[10];\frac32} = y_8^{{1\over2}}\,\bE_{3\over2}(T)- {4
   \pi\over3}\,\log y_8+O(e^{- \sqrt{y_8T_2}},e^{- \sqrt{y_8/T_2}})\,.
\end{eqnarray}
 The exponential terms in this expression are suppressed for fixed $T_2$ -- the Poisson resummation has resummed the effect of light wrapped branes and non-perturbative objects.
The net result is that  the effect of including these non-perturbative
effects has swamped the perturbative term and the leading piece is the
term     proportional     to     $y_8^{1/2}$  (and the coefficient of the
subleading logarithmic term appears with a different coefficient from
the one in the perturbative expansion discussed earlier).

\bigskip
{\bf $\bullet$ The $\partial^4\, \cR^4$  interaction in $D=7$}
\medskip

The perturbative part of  the $\calE^{(7)}_{(1,0)}$ given in last line
of~\eqref{tabd4r4}
was  derived in~\cite{Green:2010wi}.   We  are now  interested in  the
limit, $y_7\to  \infty $.  This gives  (see   (B.78)  and  (B.93)
in~\cite{Green:2010wi} with the  replacement $r^4=y_7$)
\bea 
\nn\bE^{SL(5)}_{[1000];s}   &=&y_7^{s\over  5}   \bE^{SL(4)}_{[100];s}  +
{2\pi^2 \Gamma(s-2)\over \Gamma(s)}\, \zeta(2s-4)\, y_7^{2- {4s\over 5}} 
+ O(e^{-(y_7v_3)^{\half}},e^{-(y_7\ell_s/r_i)^{\half}})\,,\\
 \nn \bE^{SL(5)}_{[0010];s} & =&y_7^{3s\over  5}\zeta(2s-1)
\bE^{SL(4)}_{[001];s}    +    {\pi    \Gamma(s-1)\over    \Gamma(s)}\,
y_7^{1-{2s\over     5}}     \bE^{SL(4)}_{[010];s-{1\over    2}}     \,
+
O(e^{-(y_7v_3)^{\half}}, e^{-(y_7\ell_s/r_i)^{\half}})\,.\\
\eea
 As before we have resummed all the instanton effects so that the exponential terms in
this expression are suppressed in the large-$y_7$ limit.
 In  particular, the series of relevance to
the  $\partial^4\,   \cR^4$  interaction  arise   in  the  combination
$\bE^{SL(5)}_{[1000];{5\over         2}+\epsilon}         $        and
$\bE^{SL(5)}_{[0010];{5\over 2}-\epsilon} $  in the limit $\epsilon\to
0$.  The poles in the individual series cancel and the combination has
the expansion, 
\begin{eqnarray}
\calE^{(7)}_{(1,0)} &=&\frac12\, \hbE^{SL(5)}_{[1000];{5\over 2}}+ {3\over \pi^3}   \hbE^{SL(5)}_{[0010];{5\over 2}} \\
\nn&=&{\pi\over  30}  y_7^{3\over  2}  \bE^{SL(4)}_{[001];{5\over  2}}  +\frac12\,
y_7^{1\over  2}  \bE^{SL(4)}_{[100];{5\over  2}}+{2\over  \pi^2}  \hat
\bE^{SL(4)}_{[010];2 } - {4\pi^2 \over 5}\log y_7+O(e^{-(y_7v_3)^{\half}}, e^{-(y_7\ell_s/r_i)^{\half}})\,.
\end{eqnarray}
 The leading
behaviour is  dominated  by the term that behaves as
$y_7^{3/2}$  (and, once more,  the coefficient of the logarithmic term is different from the one in the perturbative expansion).

These  expressions  illustrate   that  the  perturbative  supergravity
logarithms   are  dominated  by   the   ``non-perturbative''   instanton
contributions.  Furthermore, the result of summing these contributions
leads to expressions that diverge badly in the $y_D \to \infty$ limit.
This is a sign that the low energy expansion in powers of $k_i\cdot k_j\,
\ell_D^2$ is invalid.  As pointed out in~\cite{Green:2007zzb}, string
dualities  relate this  limit to  a limit  which may  be  described by
trans-Planckian scattering  in a  decompactified dual of  the original
string theory.

 \section{Comments on higher-order interactions and higher-loop supergravity.}
 \label{sec:comments} 
 
 The  structure of the  terms in  the low  energy expansion  of string
 theory that we  have discussed is presumably highly  constrained by a
 combination of  duality and  maximal supersymmetry, even  though this
 has  not been explicitly  used in  determining the  coefficients.  It
 would obviously be of interest  to discover the detailed structure of
 such  terms  and  to  what  extent  they  are  protected  by  maximal
 supersymmetry.   Although this  has  not been  understood in  detail,
 there is  some information about  higher-order terms (i.e.,  terms of
 order $\partial^8 \, \cR^4$ and higher) in $D=9$ dimensions.
 
 This  comes from  evaluating  the  amplitude in  the  limit of  large
 volume, $\calV_{(d+1)}$, of the M-theory torus (limit (iii) described
 in  the   section~\ref{sec:reviewsec})  where  the   Feynman  diagram
 approximation  to  eleven-dimensional  supergravity  compactified  on
 $\calT^{d+1}$, should be a  valid approximation. The contributions to
 the $\partial^8  \cR^4$ coefficient, $\calE^{(9)}_{(2,0)}$,  from one-loop and
 two-loop  Feynman  diagrams compactified  on  $\calT^2$  is given  in
 equation (4.25) of ~\cite{Green:2008bf}. 
This is a  {\it  sum}  of automorphic  functions  satisfying
 inhomogeneous Laplace equations with  source terms that are quadratic
 in  the  lower order  coefficients,  generalising  the equation  that
 determines             the            $\partial^6\cR^4$            coefficient
 \eqref{laplaceeigenthree}.  Although  the  expression  is  incomplete
 since  it  undoubtedly gets  contributions  from higher-loop  Feynman
 diagrams, it  is striking that its  perturbative expansion terminates
 at genus five, rather than genus four.  
 
 The occurrence  of a five-loop  contribution to $\partial^8 \cR^4$  is novel
 since  it breaks  the pattern  set by  $\partial^{2k} \cR^4$  interactions with
 $k=2,3$, for which there are  no contributions with genus larger than
 $k$ for any value of $D$.  Similar statements also apply to the other
 higher  order terms  considered  in~\cite{Green:2008bf}, namely,  the
 $\partial^{10}\cR^4$  coefficient   (equation~(4.31)  of  ~\cite{Green:2008bf})
 which  contains  terms  up  to  genus  seven,  and  the  $\partial^{12}  \cR^4$
 coefficient  (equations~(4.32) and  (4.33)  of ~\cite{Green:2008bf}),
 which includes terms  up to genus nine.  This  pattern shows that the
 claim~\cite{Berkovits:2006vc}, that supersymmetry protects $\partial^{2k}\cR^4$
 interactions  with  $1  <k\le   5$  from  renormalisation  in  $D=10$
 dimensions, must be modified in  lower dimensions.  Furthermore, there are
 indications   based  on   technical   issues  in   the  pure   spinor
 formalism~\cite{Berkovits:2009aw}  that even  in  ten dimensions  the
 non-renormalisation property  only holds up  to $k=3$.  If  that were
 the case the  $\partial^8 \cR^4$ interaction would be unprotected and would
 be  expected to  have  contributions to  all  orders in  perturbation
 theory.
  
 Following the earlier considerations of this paper, a genus-five term
 in  the  complete  $\calE^{(D)}_{(2,0)}$  coefficient would  imply  a
 five-loop logarithmic ultraviolet  divergence in maximal supergravity
 in  critical dimension $24/5$.   This contrasts  with the  value that
 follows if  the five-loop amplitude  first contributes at  order $\partial^{10}
 \cR^4$, in which  case the critical dimension would satisfy  $D_L = 4 +
 6/L$ with $L=5$, or $D_5=26/5$.   Furthermore, if $\partial^8 \cR^4$ is indeed
 not protected by supersymmetry,  so its complete coefficient contains
 terms to all orders in perturbation theory, the critical dimension at
 $L$ loops would be $D_L =  2 +14/L$.  This would lead to a seven-loop
 logarithmic ultraviolet  divergence in maximal  supergravity in $D=4$
 This  is  in  line  with  the  suggested  presence  of  a  seven-loop
 counterterm~\cite{Howe:1980th}.   This  conflicts  with  an  earlier
 argument  by the  present authors,  based on~\cite{Berkovits:2006vc},
 that  the  first divergence  would  not  occur  until at  least  nine
 loops~\cite{Green:2006yu}.

\section{Summary and discussion of higher-order contributions}
 \label{s:summary}
 
This paper has demonstrated several main features of the structure of the
duality invariant  coefficients, $\calE^{(D)}_{(p,q)}$, of  terms up
to the order $\partial^6\, \cR^4$ (or $2p+3q \le 3$) in the low-energy expansion of the four-supergraviton amplitude in type II string theory compactified to $D=10-d$ dimensions on a $d$-torus, $\calT^d$.  The explicit expressions for these coefficients were derived and their properties analysed in~\cite{Green:2010wi} (where earlier work is reviewed).

$\bullet$ The perturbation expansions of these coefficients in certain
critical dimensions -- $D_1=8$  for $\cR^4$, $D_2=7$ for $\partial^4\,
\cR^4$, $D_3=6$ for $\partial^6\, \cR^4$ -- contains logarithms of the
string coupling  $ \log y_{D_L}$.   Their presence is  required by
the duality  invariance of  the analytic part  of the  amplitude and
arises from the  presence of poles in Eisenstein  series, although the
poles themselves cancel, leaving a finite amplitude. 
Such non-analytic behaviour in the coupling constant cannot be present
in perturbative string theory so  it must disappear when the amplitude
is transformed from  the Einstein frame to the  string frame using the
relation  of the  $D$-dimensional Planck  scale to  the  string scale,
$\ell_D^{D-2} = \ell_s^{D-2}\, y_D$.  In order for this to happen there must
be specific  terms that are  logarithmic in the  Mandelstam invariants
$\sim \log(-s \ \ell_s^2/ \mu)$  (where $\mu$ is an arbitrary constant),
which  correspond to  threshold  terms in  the  amplitude.  These  are
precisely   the   threshold  $\log(-s\,   \ell_d^2)$'s   that  arise   in
supergravity  field theory  as  ultraviolet divergences,  or poles  in
dimensional regularisation.  In other  words, we have  obtained the
coefficients of the ultraviolet divergences of maximal supergravity at
$L=1$ loop in $D=8$, $L=2$ loops  in $D=7$ and $L=3$ loops in $D=6$ as
a consequence  of U-duality  rather than calculating  the supergravity
loop diagrams explicitly. 

$\bullet$ The  coefficient functions also contain  more subtle effects
associated with logarithmic  divergences in supergravity amplitudes in
dimensions $D>D_L$.  For example, we saw that the normalisation of the
double-pole, $1/\epsilon^2$, in three-loop supergravity in $D=8+2\epsilon$ dimensions is in correspondence with the coefficient of $\log^2 y_8$ in the perturbative expansion of the automorphic coefficient of the $\partial^6\cR^4$ interaction, $\calE^{(8)}_{(0,1)}$, which satisfies \eqref{laplaceeigenthree} with $D=8$.  In this case the source term on the right-hand side of \eqref{laplaceeigenthree} is the square of the $\cR^4$ coefficient, $\calE^{(8)}_{(0,0)}$, which itself has a one-loop $\log y_8$, as exhibited in \eqref{e:E322reg}.  There are plenty of further examples of logarithmic divergences in field theory in dimensions $D>D_L=4+6/L$, but they are all associated with interactions $\partial^{2L}\cR^4$ with $L>3$.

$\bullet$ The  supergravity  limit of string theory,  $\ell_s \to 0$ with $\ell_D$ fixed requires $y_D \to \infty$.  In this limit the highest-genus  perturbative term  (the highest power of $y_D$) dominates the lower-genus contributions.  However, an accumulation of an infinite number of unsuppressed instanton contributions dominates the amplitude.  These are terms that are exponentially small in the string perturbation theory limit.  The precise consequences of summing over such zero-action instanton contributions  were deduced by explicitly expanding the coefficient functions in the $y_D\to \infty$ limit.    In the cases considered here, where the torus dimension $d\le 4$, the instantons correspond (in type IIB language) to wrapped $(p,q)$-string world-sheets and $D$-instantons in $D=8$, as well as wrapped $D3$-brane world-sheets in the $D=6$ case.   One lesson to draw from this is that, as discussed in~\cite{Green:2007zzb}, supergravity cannot be decoupled from string theory\footnote{For alternative ideas on this subject see~\cite{Bianchi:2009wj}.}.

As was emphasised in section~\ref{sec:comments}, understanding the systematics of higher derivative terms is intimately related to understanding the order at which ultraviolet divergences of four-dimensional $N=8$ supergravity first arise and the stringy origin of such divergences.

 \acknowledgments
We  are  grateful  to  Stephen  Miller for  many  insights  concerning
Eisenstein series and to Jonas Bjornsson, Nick Dorey, Lance Dixon, Sergio Ferrara,
Francisco Morales, Hugh Osborn and Augusto Sagnotti for useful comments.  PV would like to thank the INFN
laboratory  at Frascati and  the LPTA  of Montpellier  for hospitality
when this work was being finalized.
MBG  is  grateful for  the  support  of  a European  Research  Council
Advanced Grant No. 247252.  J.R. acknowledges support by MCYT Research
Grant No.  FPA 2007-66665 and Generalitat de Catalunya under project 2009SGR502.

 \appendix
 \section{The $\partial^6\cR^4$ interaction in $D=6$ dimensions.}
\label{sec:d6r4}
 
 Since the coefficient $\calE^{(6)}_{(0,1)}$ was not discussed in \cite{Green:2010wi} its properties will be discussed in this appendix.
 As explained in section~\ref{sec:sugralogs}, this coefficient  satisfies the Poisson equation (\ref{laplaceeigenthree})
\be
\Delta^{(6)} \cE_{(0,1)}^{(6)} = - \Big( \bE^{SO(5,5)}_{[10000];{3\over 2}} \Big)^2 + c\,,
\label{ura}
\ee
where $c$ is a numerical constant to be determined. We have used the fact that  the coefficient of $\cR^4$ is 
$\cE_{(0,0)}^{(6)}=\bE^{SO(5,5)}_{[10000];3/ 2}$, which was discussed in detail in \cite{Green:2010wi}.

We begin by discussing  the perturbative expansion, which is associated
with  the  parabolic  subgroup  $P_{\alpha_1}$,  with  Levi  component
$GL(1)\times SO(4,4)$.  In expanding the source term in \eqref{ura} in powers of $y_6$ we need the expansion (see~(3.54) of \cite{Green:2010wi}), 
\begin{equation}
 \int_{P_{\alpha_1}}  \bE^{SO(5,5)}_{[10000];{3\over 2}} = 2\zeta(3)
    y_6^{-{3\over 2}}+ 2y_6^{-{1\over 2}}\, \bE^{SO(4,4)}_{[1000];1} \,,
   \label{perturbinsix}
\end{equation}
where the notation indicates  an integration over the instanton phases
associated  with  the  unipotent  radical, $N$,  associated  with  the
maximal parabolic subgroup $P_{\alpha_1}$, as defined in~\cite{Green:2010wi}, thereby projecting onto the zero Fourier mode.
The solution of  (\ref{ura}) can be found in perturbation theory, by expanding the automorphic function $\cE_{(0,1)}^{(6)} $  as a power series in $y_6$,
\be
\cE_{(0,1)}^{(6)}\bigg|_{\rm pert.} =y_6^{-3} \sum_{k=0}^2 y_6^k F^{SO(4,4) }_k + F_3^{SO(5,5)}\ ,
\label{ansor}
\ee
where  $  F^{SO(4,4)  }_k  $  are    perturbative  genus  $k=0,1,2$
contributions and
 \be
 \Delta^{SO(5,5)} \, F_3^{SO(5,5) }= c\ .
 \label{poch}
 \ee
We now use the decomposition of the Laplace operator (also discussed in \cite{Green:2010wi}),
\be
\Delta^{SO(5,5)}\to \Delta^{SO(4,4)} +2(y_6\partial_{y_6})^2+ 8 (y_6\partial_{y_6})\ .
\label{hhp}
\ee
Substituting (\ref{perturbinsix}), (\ref{ansor}) and (\ref{hhp}) into (\ref{ura}), we find the following equations 
\bea
&& 6F_0^{SO(4,4) } =  4\zeta(3)^2 \ ,
\non\\
&& \big(\Delta^{SO(4,4)} -8) F_1^{SO(4,4) }= -8\zeta(3) \bE^{SO(4,4)}_{[1000];1}\ ,
\\
\nn && \big(\Delta^{SO(4,4)} -6) F_2^{SO(4,4) }= -4 (\bE^{SO(4,4)}_{[1000];1})^2\ .
\label{petra}
\eea
which determine the coefficients of $F^{SO(4,4)}_k$. In particular, it
follows immediately that the tree-level and one-loop coefficients are
\begin{eqnarray}
\nn
F_0^{SO(4,4) } &=&{ 2\zeta(3)^2\over3}\ , \\
F_1^{SO(4,4)}&=& {2\zeta(3)\over3} \, \bE^{SO(4,4)}_{[1000];1}\,.  
\end{eqnarray}
The genus-two function $F_2^{(SO(4,4)}$, satisfying the last equation in \eqref{petra}, is more complicated but its properties can be analysed following the same procedure as in~\cite{Green:2005ba}, although we will not need its properties here.

We now turn to   $F^{SO(5,5) }$, which, as we will see later,  generates a logarithm that
is related to the $1/\epsilon$ pole in $D=6$ three-loop supergravity.
The most general solution of  (\ref{poch}) is a particular solution plus a solution of the homogeneous equation, where 
the homogeneous solution is  a linear combination of $SO(5,5)$ Eisenstein series.  These satisfy Laplace 
equations with eigenvalues given by~(\ref{e:LaplaEisenstein}).  The two series of relevance are  
$\bE^{SO(5,5)}_{[00010];s} $, $\bE^{SO(5,5)}_{[00001];s}$, which  satisfy 
\bea
&& \Delta^{SO(5,5)} \bE^{SO(5,5)}_{[00001];s} = {5\over 2}\, s (s-4)\ \bE^{SO(5,5)}_{[00001];s} \,,\nn
\\
&& \Delta^{SO(5,5)} \bE^{SO(5,5)}_{[00010];s} = {5\over 2}\, s (s-4)\ \bE^{SO(5,5)}_{[00010];s} \,.
\label{ochoz}
\eea
The other possible series, $\bE^{SO(5,5)}_{[00100];s}$, $\bE^{SO(5,5)}_{[01000];s}$  and $\bE^{SO(5,5)}_{[10000];s}$, need not be considered because they do not have perturbative expansions that contain powers of $y_6$ that  are consistent with string perturbation theory.   In order for \eqref{ochoz} to have zero eigenvalues as required by (\ref{poch}), we set $s=4$ (the choice $s=0$ gives  equivalent solutions).  Each series has a pole in $\epsilon$ at $s=4+\epsilon$, which needs to be subtracted, leaving an automorphic function that satisfies the Poisson equation with a constant source. The Eisenstein series with the pole subtracted will be denoted by a hat in the conventional manner.  
We are thus led to the ansatz
\begin{equation}\begin{split}
       F_3^{SO(5,5)             }           & =             a_0
\lim_{\epsilon\to0}\big(\bE^{SO(5,5)}_{[00001];4+\epsilon}        +
\bE^{SO(5,5)}_{[00010];4-\epsilon}\big)\cr
 &= a_0
\big(\hbE^{SO(5,5)}_{[00001];4}   +
\hbE^{SO(5,5)}_{[00010];4}\big)\ ,
\label{hansa}
\end{split}\end{equation}
where $a_0$ is a numerical constant discussed below.

We  are now  interested in  the constant  term of  $\hat  F^{SO(5,5) }
$ on  the parabolic subgroup $P_{\alpha_1}$,  corresponding to string
perturbation theory.  Expanding for small $y_6$ gives  an expansion of
the form 
\begin{equation}
\int_{P_{\alpha_1}} \bE^{SO(5,5)}_{[00010];s} =\pi^2\, {\zeta(2s-4)\Gamma(s-2)\over\zeta(2s)\Gamma(s)}\, y_6^{{1\over2}(s-4)} \bE^{SO(4,4)}_{[0001];s-1}+ y_6^{-{s\over 2}} \bE^{SO(4,4)}_{[0010];s}\,,
\label{hara}
\end{equation}
and the functional relation 
\begin{equation}
    \bE^{SO(5,5)}_{[00010];s} =\pi^5\,\frac{ \Gamma(s-\frac{7}{2}) \Gamma(s-\frac{5}{2}) \zeta(2s-7) \zeta(2 s-5)}{\Gamma(s-1) \Gamma(s) \zeta(2 s) \zeta(2s-2)}\,\bE^{SO(5,5)}_{[00001];4-s}
\end{equation}
and we are interested in $s\to 4$.   The first term is a genus three term which will contribute to the $\log y_6$ piece, whereas the second term is a genus one contribution that will not concern us in this discussion.

The triality symmetry of $SO(4,4)$ implies that the series
  $ \bE^{SO(4,4) }_{[1000]s} $, $ \bE^{SO(4,4) }_{[0010]s} $ and $ \bE^{SO(4,4) }_{[0001]s}$ all have
eigenvalues equal to $2s(s-3)$. Therefore, for $s=3$ these Eisenstein series solve a Laplace equation with zero eigenvalue.
{}In this case, the Eisenstein series have poles, as can be seen, for example,  from the expansion in~(C.7) of~\cite{Green:2010wi} ,  
\begin{equation}
\bE^{SO(4,4)}_{[1000];3+\epsilon}=    V_{(4)}^{ 3\over 2}  \bE^{SL(4)}_{[001];3}+
{15\over   2\pi^2}\,    \zeta(3)  \big(
{\pi^2\over \epsilon}+ \hbE^{SL(4)}_{[100];2}-{\pi^2\over4}\log V_{(4)}\big) +O(\epsilon)+n.p.
\end{equation}
where   we    have   also    used   the   $\epsilon$    expansion   of
$\bE^{SL(4)}_{[100];2+\epsilon}$ given in equation~(B.12) of~\cite{Green:2010wi}.
 The series  $\bE^{SO(4,4)}_{[0010];3+\epsilon}$, $\bE^{SO(4,4)}_{[0001];3+\epsilon}$ also have poles at $\epsilon\to 0$ with the same residue.
 
 It is now straightforward to obtain the regularised series  $\hat F^{SO(5,5)}= a_0(\hbE^{SO(5,5)}_{[00001];4} + \hbE^{SO(5,5)}_{[00010];4})$ from
$\bE^{SO(5,5)}_{[00001];4+\epsilon} + \bE^{SO(5,5)}_{[00010];4-\epsilon}$, and hence, from $ F^{SO(5,5)}$ defined by \eqref{hansa}.  Concentrating on the $\log y_6$ piece this gives 
\begin{equation}
 F_3^{SO(5,5) } 
\to {525\over 4\pi^2}\ a_0  \, \zeta(3)\, \log y_6+\cdots
\end{equation}
Finally, the value of $a_0$ can be determined by the decompactification limit $r_3\to\infty$, where we must recover the
$D=7$ genus-three automorphic functions.
One must have (see~(5.41) in \cite{Green:2010wi})
\begin{equation}
 F_3^{SO(5,5) }\to   \ 2 r_3^{3}\ \big( \bE^{SL(4)} _{[100];3}+ \bE^{SL(4)} _{[001];3}\big)\ .
\label{deco}
\end{equation}
In this limit
\be
\hbE^{SO(4,4)}_{[0010];3}+\hbE^{SO(4,4)}_{[0001];3}\to r_3^3\, \big( \bE^{SL(4)} _{[100];3}+ \bE^{SL(4)} _{[001];3}\big)+ \cdots
\ee
which requires $a_0=4\pi^2/35$. Thus
\be
 F_3^{SO(5,5) } \to 15 \zeta(3)\, \log y_6+\cdots
\ee
This means, in particular,  that
\be
c = 8 \times 15\zeta(3)
\label{cres}
\ee

\section{Normalisations}
\label{sec:normalisations}

This   appendix  gives a brief definition of the conventions used for  the
normalisations of the amplitudes.

The  normalisations  of  the  supergravity field  theory amplitude calculations at from tree level to three loops are given by~\cite{Bern:1998ug,Bern:2007hh,Bern:2008pv}
\begin{equation}\label{sugranorm}
  A^{sugra}_D=\cR^4\,\left(\kappa_{(D)}\over2\right)^2\,\left( {64\over stu\, \ell_D^6}+ \left(\kappa_{(D)}\over 2\right)^2 \, I_1 +
  \left(\kappa_{(D)}\over2\right)^4\,  I_2+
  \left(\kappa_{(D)}\over2\right)^6\, I_3+\cdots\right)\,.
\end{equation}
By convention, the Newton constant in dimension $D\le 10$, , $\kappa_D$, is related to the Planck length, $\ell_D$,  by  $2\kappa_{(D)}^2=(2\pi)^{D-3}\, \ell_D^{D-2}$.

For the purpose of comparing the field theory and string  theory  normalisations it is useful to  recall the
expansion of the tree-level amplitude string in  ten dimensions, 
\begin{eqnarray}\label{stringnorm}
\nn  A_{tree}^{string}&=&-{1\over                                     y_{D}}\cR^4\,
  {\Gamma(-{\ell_s^2s\over4})\Gamma(-{\ell_s^2t\over4})\Gamma(-{\ell_s^2u\over4})\over
    \Gamma(1+{\ell_s^2s\over4})\Gamma(1+{\ell_s^2t\over4})\Gamma(1+{\ell_s^2u\over4})}\\
&=&-{1\over             y_{D}}\cR^4\,\left({3\over           \hat\sigma_3}+2\zeta(3)+\zeta(5)\,\hat\sigma_2+{2\zeta(3)^2\over3}\, \hat\sigma_3+\cdots\right)\, .
\end{eqnarray}
where $\hat\sigma_n= (s^n+t^n+u^n)\,\ell_s^{2n}/4^n$.

\end{document}